\documentclass[12pt]{iopart}
\usepackage{iopams}

\usepackage{psfrag}
\usepackage{graphicx}
 
\newcommand{\ds}{\displaystyle}

\newcommand{\nn}{\nonumber}
\newcommand{\yp}{{y_+}}
\newcommand{\ym}{{y_-}}
\newcommand{\ypm}{{y_\pm}}

\newcommand{\qp}{{q_+}}
\newcommand{\qm}{{q_-}}
\newcommand{\qpm}{{q_\pm}}

\newcommand{\gpm}{g^\pm}

\newcommand{\Gpm}{G^\pm}

\newcommand{\Bp}{B^+}
\newcommand{\Bm}{B^-}
\newcommand{\Bpm}{B^\pm}

\newcommand{\Ap}{A^+}
\newcommand{\Am}{A^-}
\newcommand{\Apm}{A^\pm}

\newcommand{\cH}{{\cal H}}

\renewcommand{\l}{\lambda}

\renewcommand{\k}{\kappa}

\newcommand{\w}{\omega}

\begin{document}

\title{A self-interacting partially directed walk subject to a force}

\author{R.~Brak$^1$, P.~Dyke$^2$, J.~Lee$^2$, A.L.~Owczarek$^1$,\\
  T.~Prellberg$^3$, A.~Rechnitzer$^4$  and
  S.G.~Whittington$^2$}
\address {$^1$Department of Mathematics and Statistics,\\
  The University of Melbourne,\\Parkville, Victoria 3010, Australia\\
  $^2$Department of Chemistry,\\
  University of Toronto,\\Toronto, Canada, M5S 3H6.\\
  $^3$School of Mathematical Sciences,\\ 
  Queen Mary, University of London\\
  Mile End Road, London E1 4NS, UK.\\
  $^4$Department of Mathematics,\\ University of British Columbia,\\
  Vancouver,  Canada, V6K 1ZT.
}

\date{\today}


\begin{abstract}
  
  We consider a directed walk model of a homopolymer (in two
  dimensions) which is self-interacting and can undergo a collapse
  transition, subject to an applied tensile force. We review and
  interpret all the results already in the literature concerning the
  case where this force is in the preferred direction of the walk. We
  consider the force extension curves at different temperatures as
  well as the critical-force temperature curve. We demonstrate that
  this model can be analysed rigorously for all key quantities of
  interest even when there may not be explicit expressions for these
  quantities available. We show which of the techniques available can
  be extended to the full model, where the force has components in the
  preferred direction and the direction perpendicular to this. Whilst
  the solution of the generating function is available, its analysis is
  far more complicated and not all the rigorous techniques are
  available.  However, many results can be extracted including the
  location of the critical point which gives the general
  critical-force temperature curve. Lastly, we generalise the model to
  a three-dimensional analogue and show that several key properties
  can be analysed if the force is restricted to the plane of preferred
  directions.

\end{abstract}
\maketitle

\pagebreak
\section{Introduction}
\setcounter{equation}{0}

The development of atomic force microscopy 
and optical tweezers has allowed experimentalists to micro-manipulate 
individual polymer molecules (\emph{e.g.\ }Bemis \emph{et al.\ }(1999), Haupt
\emph{et al.\ }(2002), Gunari \emph{et al.\ }(2007)), 
and this has led to a considerable body
of theoretical work describing the response of a polymer to an applied 
force (\emph{e.g.\ }Halperin and Zhulina~(1991),
Cooke and Williams (2003), Rosa \emph{et al.\ }(2003)).
Several situations have been investigated, including pulling a 
polymer off a surface at which it is adsorbed, pulling a polymer from a 
preferred solvent to a less preferred solvent, pulling a copolymer which is 
localized at an interface between two immiscible liquids, and pulling
a collapsed polymer (in a poor solvent) to an extended form.  In this paper
we shall be concerned with the latter problem.

The stress-strain curve of a linear polymer in a poor solvent, being
pulled in an AFM experiment, has been measured by several groups
(Haupt \emph{et al.\ }2002, Gunari \emph{et al.\ }2007).  The
force-extension curve shows a characteristic plateau.  For forces
below this critical value the polymer will be in a collapsed state,
while for forces above the critical value the polymer will be
stretched.  The plateau region would seem to indicate a first-order
phase transition.  Indeed Grassberger and Hsu (2002) have studied
self-avoiding walks with nearest-neighbour attraction and an applied
force at low temperatures (poor solvents): they predicted a first-order
phase transition in three dimensions. On the other hand, they see no
sign of a first-order transition in two dimensions.

A well-studied exactly solved model of poor solvent polymers is the
self-interacting partially directed self-avoiding walk model (IPDSAW).
We begin our discussion by noting that a partially directed self-avoiding walk on the
square lattice without self-interaction is intrinsically anisotropic
with a preferred direction so that the polymer's size scales
proportionally to its length, and one perpendicular to this so that
the polymer's size in this direction scales sub-linearly.  As we shall
consider the square lattice with the polymer oriented one way,
the preferred direction will be the \emph{horizontal} direction and
the other direction will be the \emph{vertical} direction.  

In the absence of an applied force the critical point of this model
was found by Binder \emph{et al.\ }(1990), and is expected to model
the polymer collapse transition (or $\theta$-point). The exact
solution of the generating function was found by Brak \emph{et al.\ 
}(1992) and its singularity structure was rigorously elucidated. Prellberg
\emph{et al} (1993) used recurrence relations to generate very long
series to estimate the exponents and the scaling function for the
phase transition. A second-order phase transition similar to the
$\theta$-point was found. The tricritical nature of this transition
was described by Owczarek \emph{et al.\ }(1993) based upon small
parameter expansions and calculations of a related version of the
model (see below).

Various calculations were made by Owczarek \emph{et al.\ }(1993) that
are worth noting. Firstly a semi-continuous version of the model was
solved explicitly, building on works by Zwanzig and Lauritzen (1968
and 1970) on a related model, and showed the same tricritical nature
(identical exponents) as the lattice model (from the small parameter
expansion and the numerical work of Prellberg \emph{et al.\ }(1993)). A
later calculation by Prellberg (1995) on the asymptotics of the
generating function of the lattice model of staircase polygons
enumerated by perimeter and area implicitly demonstrated that the
scaling function and the exponents were also the same as in the fully
discrete model: the $q$-Bessel functions involved are the same in the
two lattice models. This was made explicit recently by Owczarek and
Prellberg (2007). The low temperature scaling of the partition
function was found by Owczarek (1993) in the semi-continuous model,
and once again this was the same as found by Prellberg \emph{et al}
(1993) numerically for the fully discrete case.  We argue below that
the uniform asymptotic expansion given by Owczarek and Prellberg
(2007) implies that they are indeed similar.

Secondly,  Owczarek \emph{et al.\ }(1993) generalised the fully discrete model 
to include a parameter that counts the horizontal span of
the walks. The generating function was found by generalising 
the approach of Brak
\emph{et al.\ }(1992) for the no-force case. This added parameter is
equivalent to considering a force applied in the preferred direction.
The semi-continuous model also contained this parameter. In the
semi-continuous model it was clear that this parameter did not change
the nature of the transition.  In the discrete model it was also
implicit that the collapse transition was unchanged by this parameter
(and confirmed by the work of Prellberg (1995)), and so unchanged by
such an applied force. The connection to an applied force was however
not made explicit in the paper. Rosa \emph{et al.\ }(2003) studied the
model numerically with the connection made explicit, and added
the consideration of the end-to-end distance scaling function for the
discrete model, corroborating the unchanged tricritical nature of the
transition when an applied force in the preferred direction is added.
They also plotted the critical-force-against-temperature curve for the
discrete and semi-continuous models. Again the work of Owczarek and
Prellberg (2007) makes this conclusion explicit.

Thirdly, in the appendix of Owczarek \emph{et al.\ }(1993) a further
generalisation of the model was considered and the full generating
function of this further generalisation evaluated exactly. In this
generalisation the parameter for vertical steps was replaced by two
parameters: one for steps in the positive vertical direction and one
for steps in the negative vertical direction. No analysis of this
generating function was attempted. We note that a recent paper by
Kumar and Giri (2007) considered pulling in both directions though their
focus was on finite size effects calculated numerically rather than
thermodynamic transitions and exact results.

In this paper we make explicit the connection to applying a force in
the vertical (non-preferred) direction of the generalisation discussed
above. However we go further and solve for the generating function
along the surface in the parameter space which should include the
transition point. Hence we find an explicit expression for the phase
transition point and so are able to plot exactly the critical force against
temperature curve. We also observe that one of the tricritical exponents is
unchanged when this non-preferred force is applied. This indicates,
though does not prove, that the transition may remain second-order even in
the case of this type of force. This is a little unexpected as the force
must change the end-to-end scaling of the high temperature phase when
pulling in the vertical direction from sub-linear to linear, which it
does not when applied in the horizontal direction.

We also derive functional equations for the generating function and
show how the solution of this general class of problem can be
streamlined with this approach.

Before we explain our work on the vertical pulling problem we
summarise all the known results on the horizontal pulling problem
making explicit the results in terms of the applied force. Moreover,
we apply the rigorous techniques of Brak {\it et al.\ }(1992), which had only
been applied to the case of no applied force, to the horizontal
pulling problem. In addition we discuss the behaviour of the
force-extension curves based upon the exact results. Essentially we
bring together all the known results and extend them as necessary for the
case of horizontal pulling of a partially directed polymer in two
dimensions.

At the end of this paper we consider a three-dimensional analogue of
the model and show it has similar behaviour to its two-dimensional
counterparts.


\section{Model and definitions}
\setcounter{equation}{0}
\label{model}

Consider the square lattice and a self-avoiding walk that has one end
fixed at the origin on that lattice. Now restrict the configurations
considered to self-avoiding walks such that starting at the origin
only steps in the $(1,0)$, $(0,1)$ and $(0,-1)$ directions are permitted: such a
walk is known as a partially directed self-avoiding walk (PDSAW). For
convenience, we consider walks that have 
their first step in the horizontal direction.
Let the total number of steps in the walk be $n$. We label the
vertices of the walk $i=0,1,2,\ldots , n$. Let the number of horizontal
steps be $n_x$ and the number of vertical steps be $n_y$. To define
our model we will need finer definitions, so let us define $n_{y_+}$
to be the number of $(0,1)$ steps (positive vertical steps) and
$n_{y_-}$ to be the number of $(0,-1)$ steps (negative vertical
steps). If the walk starts at the origin let the position of the other
end-point be $(s_x,s_y)$ so that the \emph{span} in the horizontal
direction is $s_x$ and the \emph{span} of the walk in the vertical
direction is $s_y$. We therefore have
\begin{eqnarray}
n &=& n_x + n_y  \nn\\
  &=& n_x + n_{y_+} + n_{y_-}
\end{eqnarray}
and 
\begin{eqnarray}
s_x & =& n_x \nn\\
s_y   &=& n_{y_+} - n_{y_-}\;.
\end{eqnarray}

\begin{figure}[ht!]
  \centering
  \includegraphics[width=12cm]{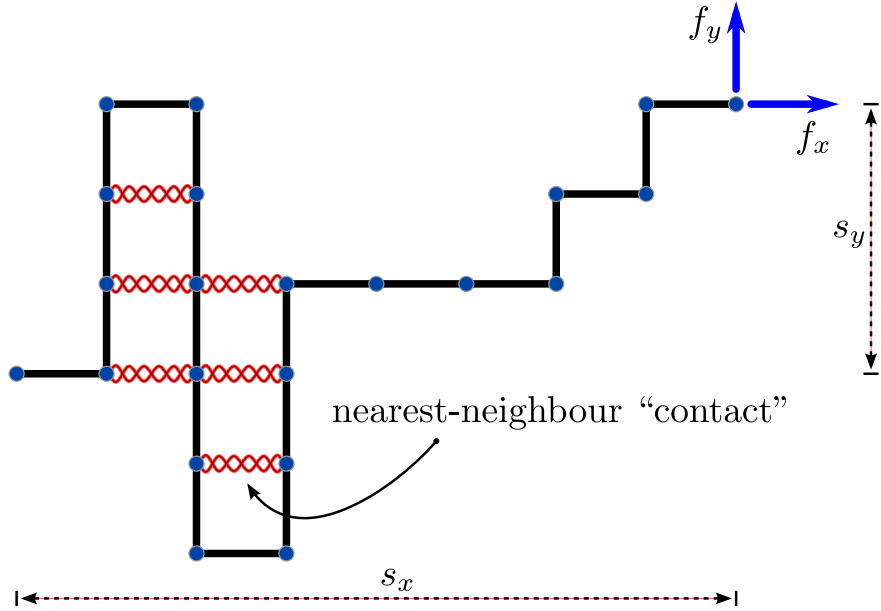}
  \caption{An example of a partially directed walk (the bold black
    path) of length $n=21$ with $n_x=8$, $n_{y_+} = 8$, and $n_{y_-} =
    5$ and having six nearest-neighbour
    `contacts' (shown as intertwined (red) curves) so $m=6$. The
    horizontal span is $s_x=8$ while the vertical span is $s_y=3$. One
    end is fixed at the origin while forces are applied to the other
    end (horizontal~$f_x$ and vertical~$f_y$).}
\label{fig-model}
\end{figure}

An example
configuration along with the associated variables of our model is 
illustrated in figure~\ref{fig-model}.
 
To define our model we add various energies and hence
Boltzmann weights to the walk. First, any two occupied sites of the
walk not adjacent in the walk though adjacent on the lattice are
denoted \emph{nearest-neighbour contacts} or \emph{contacts}: see
figure~\ref{fig-model}.  An energy $-J$ is added for each
such contact. We define a Boltzmann weight $\omega=e^{\beta J}$
associated with these contacts, where $\beta = 1/k_BT$ and $T$ is the
absolute temperature. Without loss of generality we shall take the
units of energy to be such that $J=1$ and therefore $\omega=e^\beta$,
except when we discuss pulling polymers without any
self-interaction ($\omega=1$) where we will have $J=0$. 

An external horizontal force $f_x$ pulling at the
other end of the walk adds a Boltzmann weight $h^{s_x}$ with 
$h=e^{\beta f_x}$. 
An external vertical force $f_y$ pulling at the
other end of the walk adds a Boltzmann weight $v^{s_y}$ 
with $v=e^{\beta f_y}$.

The partition function $Z_n(f_x,f_y,\beta)$ of the model for walks of
length $n\geq 1$, where for later mathematical convenience the
first step of the walk is a horizontal step, is
\begin{equation}
Z_n(f_x,f_y,\beta) =
\sum_{\varphi \mathrm{\; is\; PDSAW\; of\; length\;} n} h^{s_x(\varphi)}
v^{s_y(\varphi)}\omega^{m(\varphi)} \,,
\end{equation}
where $m(\varphi)$ is the number of nearest-neighbour contacts in the PDSAW,
$\varphi$.  The
generating function $\hat{G}(z;h,v,\omega)$ is
\begin{equation}
\hat{G}(z;h,v,\omega) = \sum_{n=1}^{\infty} Z_n(f_x,f_y,\beta) z^n \,,
\end{equation}
so $z$ can be considered as a fugacity for the steps of the walk and the
generating function as a ``generalised partition function''. We shall 
denote the radius of convergence of $\hat{G}(z;h,v,\omega)$ as a
function of $z$ as $z_c(h,v,\omega)$.
The mean values of $n$, $m$, $s_x$ and $s_y$,  
are given by
\begin{eqnarray}
\langle n \rangle &=& z\frac{\partial \log \hat{G}}{\partial z}, \quad 
\langle m \rangle = \omega\frac{\partial \log \hat{G}}{\partial
  \omega}\;,\nonumber \\
\langle s_x \rangle &=& h\frac{\partial \log \hat{G}}{\partial h}, \quad 
\langle s_y \rangle = v\frac{\partial \log \hat{G}}{\partial v}.
\end{eqnarray}

We shall define the number of $n$-edge partially directed walks with $m$
nearest-neighbour contacts, horizontal span $s_x$ and vertical span
$s_y$ as $d_n(s_x,s_y,m)$ so that 
\begin{equation}
Z_n(f_x,f_y,\beta) = \sum_{s_x,s_y,m} d_n(s_x,s_y,m) h^{s_x} v^{s_y} \omega^m \;.
\end{equation}
When $v=1$, that is $f_y=0$, we let
\begin{equation}
b_n(s_x,m) = \sum_{s_y} d_n(s_x,s_y,m)\;,
\end{equation}
and so $b_n(s_x,m)$ is the number of 
$n$-edge partially directed walks with $m$
nearest-neighbour contacts and horizontal span $s_x$. So 
\begin{equation}
Z_n(f_x,0,\beta) = \sum_{s_x,m} b_n(s_x,m) h^{s_x} \omega^m \;.
\end{equation}
Let the number of partially directed walks of length $n$ be
$\bar{b}_n$ so that
 \begin{equation}
\bar{b}_n = \sum_{s_x,s_y,m} d_n(s_x,s_y,m) = \sum_{s_x,m} b_n(s_x,m)\;.
\end{equation}

It is advantageous when working with the generating function to define
different variables. Let the generating function $G(x,\yp,\ym,\omega)$
be defined as
\begin{equation}
G(x,\yp,\ym,\omega) =
\sum_{n=1}^{\infty}
\sum_{\varphi}
x^{n_x(\varphi)} \yp^{n_{y_+(\varphi)}} \ym^{n_{y_-(\varphi)}}
\omega^{m(\varphi)}\,,
\end{equation}
where the sum over $\varphi$ is over all PDSAW's of length $n$.
Then making the substitutions
\begin{eqnarray}
x& =& hz \nn \\
y_+ &=& v z \nn \\
y_- &=& z/v
\label{substitutions}
\end{eqnarray}
demonstrates that
\begin{equation}
G(hz,zv,z/v,\omega) = \hat{G}(z;h,v,\omega).
\end{equation}
Hence we have 
\begin{eqnarray}
Z_n(f_x,f_y,\beta)&=[z^n]G(hz,zv,z/v,\omega)\nonumber\\
&={\frac{1}{2\pi i}}\oint G(hz,zv,z/v,\omega){\frac{dz}
  {z^{n+1}}}.
\end{eqnarray}
It will be useful to define
\begin{equation}
q_+ = y_+\omega \quad \mbox{ and } q_-=y_-\omega
\end{equation}
and 
\begin{equation}
q = \sqrt{q_+q_-}\;,
\end{equation}
and on making the substitutions (\ref{substitutions})
\begin{equation}
q = \omega z\;.
\label{qdef}
\end{equation}
We define the reduced limiting free energy as
\begin{equation}
\kappa(f_x,f_y,\beta) =\lim_{n\rightarrow\infty} \frac{1}{n} \log
\left[Z_n(f_x,f_y,\beta)\right]\;
\end{equation}
where the existence of the limit can be established by concatention arguments.
The radius of convergence $z_c(h,v,\omega)$ can be related to the free
energy $\kappa(f_x,f_y,\beta)$ as 
\begin{equation}
\kappa(f_x,f_y,\beta) = -\log z_c(h,v,\omega)\;.
\end{equation}

It will turn out that there is a single phase transition where the
free energy is singular as a function of $\beta$. We shall denote the
phase-transition inverse-temperature as $\beta^t\equiv\beta^t(f_x,f_y)$. In
general a superscript of $t$ implies the critical value of a
parameter, \emph{e.g.\ }the critical horizontal force at a fixed temperature
and zero vertical force is $f_x^t\equiv f_x^t(0,\beta)$. However, we
use a subscript $t$ to denote the critical values of the fugacities.  The
value of the radius of convergence of $\hat{G}(z;h,v,\omega)$ as a
function of $z$ at $\omega_t= e^{\beta^t}$ is $z_c^t(h,v)=
z_c(h,v,\omega_t(h,v))$.

When $f_y=0$ we shall set $y=\yp=\ym$ and so the generating function
we need to consider for the case where there is no vertical stretching
force is $G(x,y,y,\omega)$, noting  that $G(hz,z,z,\omega) =
\hat{G}(z;h,1,\omega)$, where
\begin{equation}
\hat{G}(z;h,1,\omega) = \sum_{n=1}^{\infty} Z_n(f_x,0,\beta) z^n =
\sum_{n,s_x,m} b_n(s_x,m) h^{s_x} \omega^m z^n\,.
\end{equation}
Note that for  $f_y=0$ we have 
\begin{equation}
q = y\omega\;,
\end{equation}
which agrees with (\ref{qdef}) when the substitutions
(\ref{substitutions}) are made since $y=z$ when $y=\yp=\ym$ (that is,  $v=1$).

In subsequent calculations it is convenient to define the generating
function $G_r(x,\yp,\ym,\omega)$ for walks that have $r$ 
vertical steps immediately after
the first horizontal step, where $-\infty < r < \infty$. It will also
be convenient to introduce the generating function
$G_r^+(x,\yp,\ym,\omega)= G_r(x,\yp,\ym,\omega)$ for walks that have
$r\geq0$ steps after the first horizontal step in the positive $y$
direction and $G_r^-(x,\yp,\ym,\omega)=G_{-r}(x,\yp,\ym,\omega)$ that
have $r\geq0$ steps after the first horizontal step in the negative
$y$ direction. Clearly, $G_0^+= G_0^-$. 

In the functional equation section (section~\ref{func-eqns-vert}) we
need
\begin{equation}
F(p) = \sum_{r=-\infty}^{\infty} G_r(x,y,y,\omega) p^r\,,
\label{Fp}
\end{equation}
\begin{equation}
F^{+}(p) = \sum_{r=0}^{\infty} G_r^{+}(x,\yp,\ym,\omega) p^r
\end{equation}
and
\begin{equation}
F^{-}(p) = \sum_{r=1}^{\infty} G_r^{-}(x,\yp,\ym,\omega) p^r\;.
\end{equation}
Note the asymmetry of the summation index in the final two definitions.
Note that 
\begin{equation}
G(x,y,y,\omega) = F(1)= \sum_{r=-\infty}^{\infty} G_r(x,y,\omega)
\end{equation} and 
\begin{equation}
G(x,\yp,\ym,\omega) = F^+(1) + F^-(1) = \sum_{r=-\infty}^{\infty}
G_r(x,\yp,\ym,\omega)  \;.
\end{equation}

\section{Pulling in the preferred direction ($f_y=0$)}
\setcounter{equation}{0}
\label{x-pull}

In this section we will begin discussing the case of horizontal pulling
force where $f_y=0$.

\subsection{No self-interactions ($\omega=1$)}
\label{x-pull-no-int}

Let us begin by considering the case of no interactions and no force so
that $J=f_x=f_y=0$.
It is easy to see that 
$\lim_{n\to\infty} n^{-1} \log \bar{b}_n=\log(1+\sqrt{2})$.  
Let us consider the case of no vertical force so that $f_y=0$ and
$v=1$. As discussed in the previous section we are interested in the
generating function $\hat{G}(z;h,1,\omega)= G(hz,z,z,\omega)$. As defined above this generating
function converges when $z < z_c(h,1,\omega)$. Recall that the free
energy $\kappa(f_x,0,\beta)$ is given by $\kappa = -\log z_c$ (assuming 
that the limit
defining the free energy exists, which will be proved in a later section).

\begin{figure}[ht!]  
  \psfrag{b}{\begin{large}$\beta f_x$\end{large}}
  \psfrag{l}[b][][1][90]{\begin{large}$\ds \lim_{z\to z_c} 
  \frac{\langle s_x\rangle}{\langle n \rangle} $\end{large}}
  \centering
  \includegraphics[width=9cm]{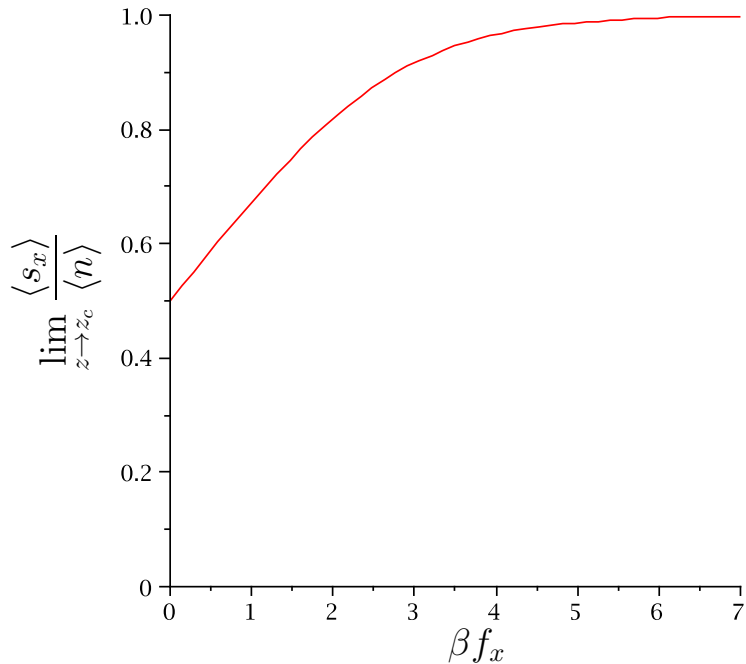}
  \caption{The dependence of $\ds \lim_{z \to z_c} \langle
  s_x\rangle /\langle n \rangle$ on $\beta f_x$ when $\omega=1$.}
\label{fig ss w1}  
\end{figure}

When $\omega=1$ (which corresponds to turning off the vertex-vertex 
interaction and hence to good solvent conditions), $\hat{G}(z;1,h,1)$
satisfies the equation
\begin{equation}
\hat{G}(z;h,1,1) =  hz(1+ \hat{G}(z;h,1,1)) + \frac{2hz^2}{1-z}(1  + \hat{G}(z;h,1,1))\;,
\end{equation}
so that
\begin{equation}
1+ \hat{G}(z;h,1,1) = \frac{ 1- z}{1-z-hz-hz^2}.
\end{equation}
From this we can readily calculate the ratio 
$\langle s_x\rangle /\langle n\rangle$ and take the 
thermodynamic limit by letting $z\to z_c(h,1,1)$
where
\begin{equation}
z_c(h,1,1)=\frac{\sqrt{h^2+6h+1}-1-h}{2h}.
\end{equation}
In figure~\ref{fig ss w1} is a plot of $\lim_{z \to z_c} \langle
s_x\rangle /\langle n \rangle$ against $\beta f_x$.  Note that the
stress-strain curve is qualitatively the same as that found
experimentally for polystyrene in a good solvent (toluene) (Gunari
\emph{et al.\ }2007, figure 1).

For $\omega \neq 1$ the situation is more difficult and we 
derive some results about the generating function in the following sections.

\subsection{Convexity and continuity for $\omega \neq 1$}
\label{convex}

In this section we establish the existence of the 
thermodynamic limit for the canonical problem, 
and prove convexity and continuity.  This implies continuity 
of the phase boundary.  

It will be convenient to consider a subset of the walks counted by
$\hat{G}(z;h,1,\omega)$.  Let $a_n(s_x,m)$ be the number of partially directed walks
with $n$ edges, $m$ contacts and $x$-span equal to $s_x$ which satisfy
the additional constraint that their last step is in the
east direction.  (That is, both the first and last steps are east steps.)
These walks can be concatenated by identifying the last vertex of 
one walk with the first vertex of the other walk which yields the inequality
\begin{equation}
a_{n_1+n_2}(s_x,m) \ge \sum_{s_{x_1},m_1} a_{n_1}(s_{x_1},m_1) a_{n_2}
(s_x-s_{x_1},m-m_1).
\end{equation}
Defining the partition function
\begin{equation}
A_n(h,\omega)=\sum_{s_x,m}a_n(s_x,m) h^{s_x}\omega^m\;,
\end{equation}
this implies the super-multiplicative inequality
\begin{equation}
A_{n_1+n_2}(h,\omega) \ge A_{n_1}(h,\omega)A_{n_2}(h,\omega).
\end{equation}
Since $A_n(h,\omega) \le 3^n \omega^n h^n$, it follows that $n^{-1} \log A_n(h,\omega)$ is bounded
above 
for $h,\omega < \infty$.  Hence the super-multiplicative 
inequality above implies the existence of the limit
\begin{equation}
\lim_{n\to\infty}n^{-1} \log A_n(h,\omega) \equiv \tilde{\kappa}(f_x,\beta)
\end{equation}
for $h,\omega < \infty$.
 
We now recall the definition of the partition function
\begin{equation}
Z_n(f_x,0,\beta) = \sum_{m,s_x}b_n(m,s_x)\omega^m h^{s_x}
\end{equation}
and define $B_n(h,1,\omega)= Z_n(f_x,0,\beta)$.
Since $A_{n+1}(h,\omega) = hB_n(h,\omega)$ it follows that 
\begin{equation}
\lim_{n\to\infty}n^{-1} \log B_n(h,\omega) = \tilde{\kappa}(f_x,\beta)=\kappa(f_x,0,\beta)
\end{equation}
for $h,\omega < \infty$.

To prove that $\kappa(f_x,\beta)$ is a convex function of $\beta$ and
$\beta f_x$
we note that H\"{o}lder's inequality implies that
\begin{eqnarray}
\left(\sum_{m,s_x}b_n(m,s_x)\omega_1^m h_1^{s_x}\right)& 
&\left(\sum_{m,s_x}b_n(m,s_x) \omega_2^m h_2^{s_x}\right)
 \nonumber\\
&\ge&\left(\sum_{m,s_x} b_n(m,s_x)(\sqrt{\omega_1\omega_2})^m
  (\sqrt{h_1h_2})^{s_x} \right)^2
\end{eqnarray}
so that 
\begin{equation}
B_n(h_1,\omega_1) B_n(h_2,\omega_2) \ge B_n(\sqrt{\omega_1\omega_2},\sqrt{h_1h_2})^2
\end{equation}
and hence
\begin{equation}
\frac{n^{-1} \log B_n(h_1,\omega_1) + n^{-1} \log B_n(h_2,\omega_2)}{2}
\ge n^{-1} \log B_n(\sqrt{h_1h_2},\sqrt{\omega_1\omega_2}).
\end{equation}
This shows that $n^{-1} \log B_n(h,\omega)$ is a convex function of $\beta$ and 
$\beta f_x$.  (It is convex as a surface, not just separately convex in each 
variable.)  Since the limit of a sequence of convex 
functions (when it exists) is a convex function, $\kappa(f_x,0,\beta)$ 
is a convex function of $\beta$ and $\beta f_x$.  Hence $\kappa(f_x,0,\beta)$
is continuous and is differentiable almost everywhere.  Since 
$\kappa(f_x,0,\beta) = -\log z_c(h,1,\omega)$, the singularity surface $z=z_c(h,1,\omega)$ is also 
continuous and differentiable almost everywhere.

\subsection{The generating function for $\omega \neq 1$}
\label{genfun}

While the calculation of the generating function for $\omega\neq 1$
appears in Owczarek {\it et al.\ }(1993) (see section 4 of that paper) we feel
that it is worth summarizing and presenting in a slightly different way.
In order to consider the case $\omega \neq 1$ it was necessary to
generalize a method originally due to Temperley (1956) and used for
the case $h=1$ by Brak \emph{et al.\ }(1992).  
%

\begin{figure}[ht!]
  \centering
  \includegraphics[width=15cm]{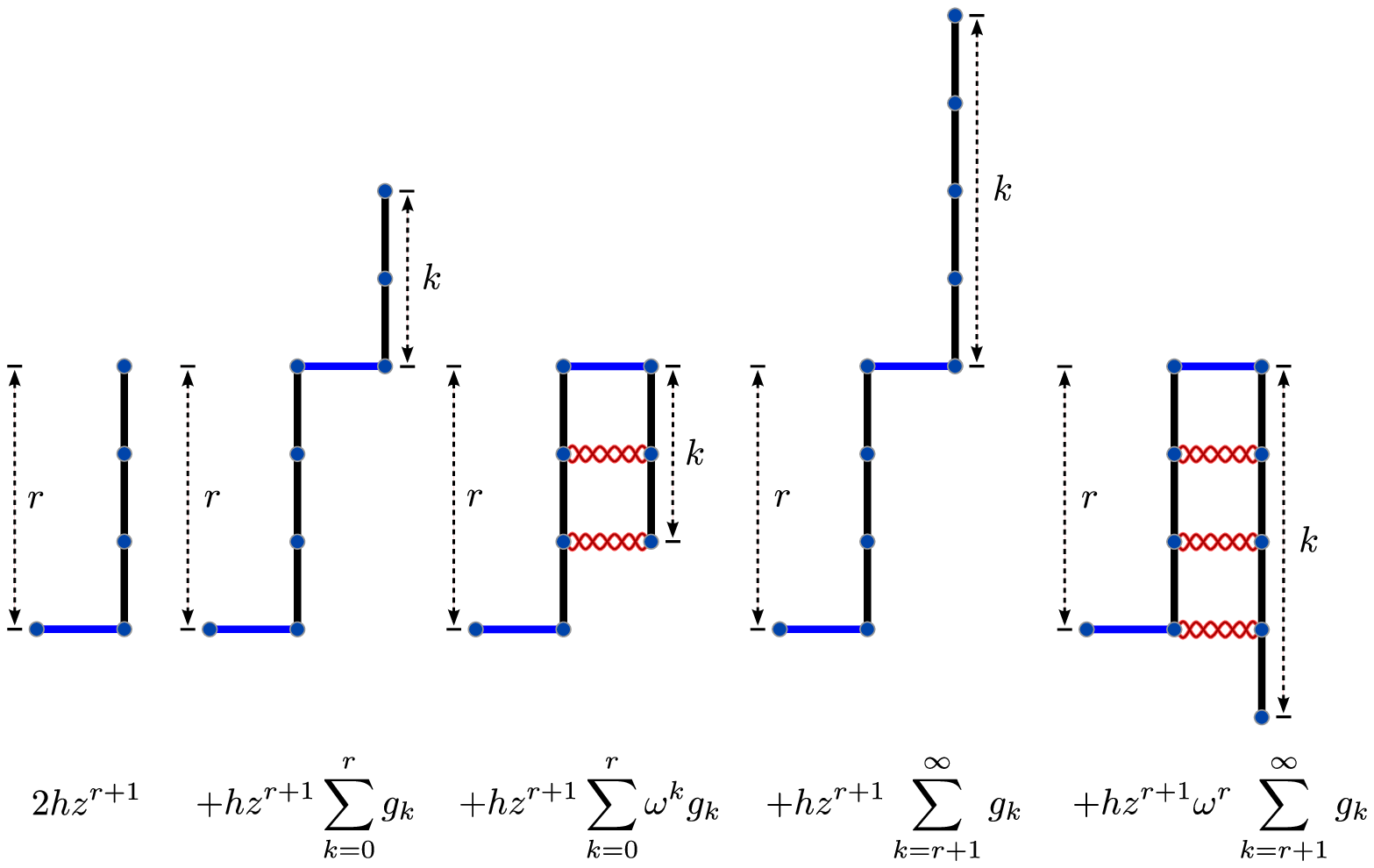}
  \caption{Walks counted by $g_r$, with $r>0$, either contain a single
horizontal bond (and so are an $\rfloor$ or an $\rceil$) or can be constructed
by appending an $\rfloor$ or $\rceil$ configuration of bonds to a walk counted
by $g_k$. Summing over all these possibilities gives equation~(\ref{sum-rec}).}
\label{fig rec gen w}
\end{figure}

For convenience let the partial generating function
$\hat{G}_r(z;h,1,\omega)$ in the case where $v=1$ (no vertical force)
be denoted as
\begin{equation}
g_r= 2 \hat{G}_r(z;h,1,\omega) \qquad r \ge 1,
\end{equation}
and $g_0=\hat{G}_0(z;h,1,\omega)$.
Now the generating function required is $\hat{G}(z;h,1,\omega)$ which is the sum over all $r$ as 
\begin{equation}
\hat{G}(z;h,1,\omega) =\sum_{r=-\infty}^{\infty}\hat{G}_r(z;h,1,\omega)\;.
\end{equation}
Noting that $\hat{G}_{-r}(z;h,1,\omega)=\hat{G}_r(z;h,1,\omega)$, we have
the partial generating functions $g_r$ satisfying the relations
\begin{equation}
g_0 = hz + hz(g_0+g_1+ \ldots) = hz(1+ \hat{G}(z;h,1,\omega))
\end{equation}
and
\begin{equation}
g_r = hz^{r+1}\left(2+ \sum_{k=0}^{r}(1+\omega^k)g_k + 
(1+\omega^r)\sum_{k=r+1}^{\infty}g_k\right)\;, 
\quad r \ge 1.
\label{sum-rec}
\end{equation}
See figure~\ref{fig rec gen w}.
These results imply that $g_r$ satisfies the recurrence
\begin{equation}
g_{r+1} - (z+\omega z)g_r + \omega^r h z^{r+2}(\omega-1)g_r + \omega
z^2 g_{r-1}=0\;.
\label{recurrence1}
\end{equation}
We have $q=\omega z$ so that (\ref{recurrence1}) becomes
\begin{equation}
g_{r+1} - (z+q)g_r +q^r h z(q-z)g_r +qzg_{r-1} = 0\;.
\label{recurrence2}
\end{equation}
This can be solved with the Ansatz
\begin{equation}
g_r = \lambda^r\sum_{m=0}^{\infty} p_m(q)q^{mr}
\end{equation}
which is a solution provided that
\begin{equation}
\lambda^2-\lambda (z+q) + qz =0
\label{indicialeqn}
\end{equation}
and
\begin{equation}
p_m(q) = \frac{\lambda hz(z-q)q^m}{(\lambda q^{m}-q)(\lambda q^m-z)}p_{m-1}(q).
\end{equation}
Since $p_0(q)=1$ this gives
\begin{equation}
p_m(q)= \frac {\lambda^m h^mz^m(z-q)^mq^{m(m+1)/2}}
{\prod_{k=1}^{m} (\lambda q^k-q)\prod_{k=1}^{m}(\lambda 
q^k-z)}.
\end{equation}

The quadratic equation (\ref{indicialeqn}) has the two solutions
$\lambda_1=z$ and $\lambda_2=q=\omega z$ and the general solution
for $g_r$, $r > 0$ is
\begin{equation}
g_r = A_1 g_r^{(1)} + A_2 g_r^{(2)}
\end{equation}
where
\begin{equation}
g_r^{(i)} = \lambda_i^r + \lambda_i^{r} \sum_{m=1}^{\infty}
 \frac {\lambda_i^m h^mz^m(z-q)^mq^{m(m+1)/2}}
{\prod_{k=1}^{m} (\lambda_i q^k-q)\prod_{k=1}^{m}(\lambda_i 
q^k-z)} q^{mr}.
\end{equation}
Arguments similar to those in Brak \emph{et al.\ }(1992) show that 
$A_2=0$ and $A_1$ can be determined by noting that
\begin{equation}
g_0 = \frac{1}{2}A_1 g_0^{(1)} = hz + hz\hat{G}(z;h,1,\omega)
\end{equation}
and 
\begin{equation}
g_1 = A_1 g_1^{(1)} = a + b\hat{G}(z;h,1,\omega)
\end{equation}
where $a=hz^2(2+hz-\omega hz)$ and $b=hz^2(1+hz+\omega -\omega hz)$.
These simultaneous equations give
\begin{equation}
\hat{G}(z;h,1,\omega)= \frac{2hzg_1^{(1)}-ag_0^{(1)}}{bg_0^{(1)}-2hzg_1^{(1)}}\;,
\label{horiz-gf-soln1-orig}
\end{equation}
which can be written as
\begin{eqnarray}
1 + \hat{G}(z;h,1,\omega) 
& =& \frac{2hz^2(\omega-1)g_0^{(1)}}{bg_0^{(1)}-2hzg_1^{(1)}}\nonumber\\
& = & \frac{(\omega-1)g_0^{(1)}}{ z (1+hz+\omega -\omega hz)g_0^{(1)}- 2 g_1^{(1)}}\;.
\label{horiz-gf-soln1}
\end{eqnarray}
The solution given by Owczarek {\it et al.\ }(1993)  was written as
 \begin{equation} 
1+G(x,y,y,\w)=\frac{1-\w}{2(\cH(y,y\w,xy(\w-1))-1)+(1-\w)(1-x)}\;.
\label{horiz-gf-soln1-opb}
\end{equation}
where
\begin{equation}
\cH(y,q,t)=\frac{H(y,q,qt)}{H(y,q,t)}
\end{equation}
and 
\begin{equation}
H(y,q,t)=\sum_{n=0}^\infty\frac{q^{{n}\choose{2}}(-t)^n}{(y;q)_n(q;q)_n}\;.
\label{Hfunction}
\end{equation}
The two forms can be seen to be the same when one notes that
\begin{eqnarray} 
1+  \hat{G}(z;h,1,\omega) 
& = & 1+G(hz,z,z,\w)\nonumber \\
& = & \frac{1-\w}{2(\cH(z,z\w,hz^2(\w-1))-1)+(1-\w)(1-hz)}\nonumber\\
& = & \frac{\w-1}{(1+hz+\omega -\omega hz) -2\cH(z,z\w,hz^2(\w-1))}
\label{horiz-gf-soln-coor}
\end{eqnarray}
and that
\begin{equation}
\cH(z,z\w,hz^2(\w-1))= \frac{g_1^{(1)}}{zg_0^{(1)}}\;.
\end{equation}

The implications for the singularity diagram of this expression for
$\hat{G}(z;h,1,\omega)$, and the phase transitions, will be considered
in the next section.

\subsection{Solution on the special surface $z=1/\omega$}

Before we proceed to discuss the singularity diagram let us consider
the solution on the special surface defined by $q=1$ since the
solution described by (\ref{horiz-gf-soln1}) is singular when
$z=1/\omega$ though the generating function is finite for large enough
$\omega$.

The recurrence for the partial generating functions becomes
\begin{equation}
g_{r+1} +[h z(1-z)-(1+z)]g_r +zg_{r-1} = 0\;,
\end{equation}
and now the simpler Ansatz of $g_r=C \mu^r$ can be used to find
\begin{equation}
1 +  \hat{G}(1/\omega; h,1,\omega) = \sqrt{\left(
     \frac{w^2(w-1)}{w(w-h)^2-(w+h)^2}\right)}\;.
\end{equation}
When $h=1$ we recover the previously calculated result
of Brak \emph{et al.\ }(1992).


\section{The phase diagram for a horizontal force }
\setcounter{equation}{0}
\label{pd_x-pull}

In order to analyse the phase transition structure we need to analyse the form of the 
singularity diagram as a function of the 
$\omega$ and $h$ parameters. We derive a functional equation for a slight
generalization of the quantity $g_r^{(1)}$.  We introduce a 
parameter $t$ and define the function
\begin{equation}
g(t;q,h,\omega) = 1+ \sum_{m=1}^{\infty}
\frac{h^m\omega ^{-m}\left(\frac{1}{\omega }-1\right)q^{m(m+5)/2}t^m}
{\prod_{k=1}^m(1-q^k)(1-q^k/\omega )}
\end{equation}
Note that $g_1^{(1)}$ is equal to $zg(tq;q,h,\omega)$ evaluated at 
$t=1$ while $g_0^{(1)}=g(1;q,h,\omega)$.  

The function $g(t;q,h,\omega)$ satisfies the functional equation
\begin{eqnarray}
g(t;q,h,\omega) +& g(q^2t;q,h,\omega)/\omega\nonumber \\  
&= 
(1+1/\omega  +(1/\omega  - 1)hq^2t/\omega )g(qt;q,h,\omega).
\end{eqnarray}
Defining
\begin{equation}
\hat{H}(t;q,h,\omega) = g(t;q,h,\omega)/g(tq;q,h,\omega)
\end{equation}
gives 
\begin{equation}
\hat{H}(t;q,h,\omega) = (1+1/\omega  +(1/\omega  - 1)hq^2t/\omega )
-\frac{1/\omega }{\hat{H}(qt;q,h,\omega)}\;,
\label{CF1}
\end{equation}
which leads to a continued fraction representation for 
$\hat{H}(t;q,h,\omega)$.   Note that
\begin{equation}
\hat{H}(1;q,h,\omega)\equiv \hat{H}(q,h,\omega) = zg_0^{(1)}/g_1^{(1)}
= \frac{1}{\cH(z,z\w,hz^2(\w-1))}
\end{equation}
and therefore 
\begin{equation}
1 + \hat{G}(z;h,1,\omega)= \frac{(\omega-1)\hat{H}(q,h,\omega)}{(1+hz+\omega -\omega hz)\hat{H}(q,h,\omega)-2}.
\end{equation}

The function $\hat{H}(q,h,\omega)$ is singular on the hyperbola $q=\omega z=1$ for all values
of $h$ and the only other singularities in $G$ come from the poles 
corresponding  to the denominator of $G$ being zero, \emph{i.e.\ }determined by the solution of the equation
$(1+hz+\omega -\omega hz)\hat{H}(q,h,\omega)-2=0$.  This line of poles intersects the 
hyperbola at a point determined by solving the equation 
$(1+hz+\omega -\omega hz)\hat{H}(1,h,\omega)-2=0$, where $\hat{H}(1,h,\omega)$ is determined by the quadratic 
equation
\begin{equation}
\hat{H}(1,h,\omega)^2-\left[1+\frac{1}{\omega}+\left(\frac{1}{\omega}-1\right)\frac{h}{\omega}\right]
\hat{H}(1,h,\omega)+\frac{1}{\omega }=0\;,
\end{equation}
which comes from (\ref{CF1}) on setting $q=t=1$.  
The variables $\omega $ and $h$ are related through the equation
\begin{equation}a
h^2(\omega -1)-2\omega (\omega +1)h +\omega ^3-\omega ^2=0
\end{equation}
which implies that the critical value of $h$ is given by
\begin{equation}
h_t=\frac{(\omega +1-2\sqrt{\omega })\omega }{\omega -1}.
\end{equation}

Let us consider $q_c(h,1,\omega)=z_c(h,1,\omega)\omega$. We have
that $0< q_c<1$ for $1\leq \omega< \omega_t(h,1)$. We have argued that there is a pole
in the generating function which means that there is a zero of 
$D=1/(1+G)$. That is, the function
\begin{equation}
D(q,\omega,h)\equiv \frac{\left[(1+hq/\omega + \omega -hq) -2\cH(q/\omega, q, h q^2(\w-1)/\w^2)\right]}{\omega-1}\;
\end{equation}
obeys
\begin{equation}
D(q_c,\omega,h)= 0\,.
\end{equation}
We also have analyticity of $\cH$ for
$0< q<1$, shown by the ratio test. When $\omega=1$ there is a simple
pole in the generating function. We have
\begin{equation}
D(q,1,h)\equiv  \frac{1- (1+h) q -h q^2}{1-q} 
\end{equation}
and so that 
\begin{equation}
\frac{\partial D(q,1,h)}{\partial q} < 0
\end{equation}
at $q=q_c(h,1,1)$. The derivative is negative for all $0<q<1$.


We now prove that there is a simple pole for all $1\leq\omega<\omega_t(h,1)$. For this, we compute
\begin{equation}
\frac{\partial D(q,\omega,h)}{\partial q}=-\frac h\omega-\frac2{\omega-1}\frac d{dq}\cH(q/\omega, q, h q^2(\w-1)/\w^2)\;.
\end{equation}
Observe that $\cH(y,q,t)$ is related to the generating function $G_{sc}(x,y,q)$ of staircase polygons enumerated by width, height, and area. Equations (4.6) and (4.9) of Prellberg and Brak (1995) imply that
\begin{equation}
G_{sc}(x,y,q)=y\left[\cH(qy,q,qx)-1\right]\;.
\end{equation}
It follows that $\cH(q/\omega, q, h q^2(\w-1)/\w^2)$ is a power-series in $q$ with non-negative coefficients. Therefore, its derivative with respect
to $q$ is positive. Hence 
\begin{equation}
\frac{\partial D(q,\omega,h)}{\partial q} < 0
\label{diffq-cond}
\end{equation}
and therefore there is a simple pole in the generating function at all values of 
$1\leq \omega<\omega_t(h,1)$.

Now, using a version of the \emph{implicit function theorem} that
gives analyticity (see chapter 6 of Krantz and Parks 2002) it follows that 
$q_c(h,1,\omega)$ is an analytic function of both $\omega$
and $h$.

To determine the detailed shape of the part of the 
phase boundary determined by the poles of $\hat{G}$ we solve the equation
$ (1+hz+\omega -\omega hz) \hat{H} -2 =0$ numerically by evaluating $H$ from its continued fraction
expansion.  That is, we write $\hat{H}$ as 
\begin{equation}
\hat{H}(t;\omega,h,q)=
(1+1/\omega +(1/\omega-1)hq^2t/\omega )(1+\beta_0 \mathcal{C})
\end{equation}
where
\begin{equation}
\mathcal{C}=    \frac{1}{\displaystyle 1
              + \frac{\beta_1}{\displaystyle 1
              + \frac{\beta_2}{\displaystyle 1
              + \frac{\beta_3}{\displaystyle \ddots }}}}
\end{equation}
and 
\begin{equation}
\beta_{k}= \frac{-1/\omega }
{(1+1/\omega +(1/\omega -1)hq^{k+2}t/\omega)
(1+1/\omega +(1/\omega -1)hq^{k+3}t/\omega )}\;.
\end{equation}

\begin{figure}[ht!]
  \psfrag{z}{\begin{Large}$z_c$\end{Large}}
  \psfrag{w}{\begin{Large}$\omega$\end{Large}}
  \centering  
  \includegraphics[width=9cm]{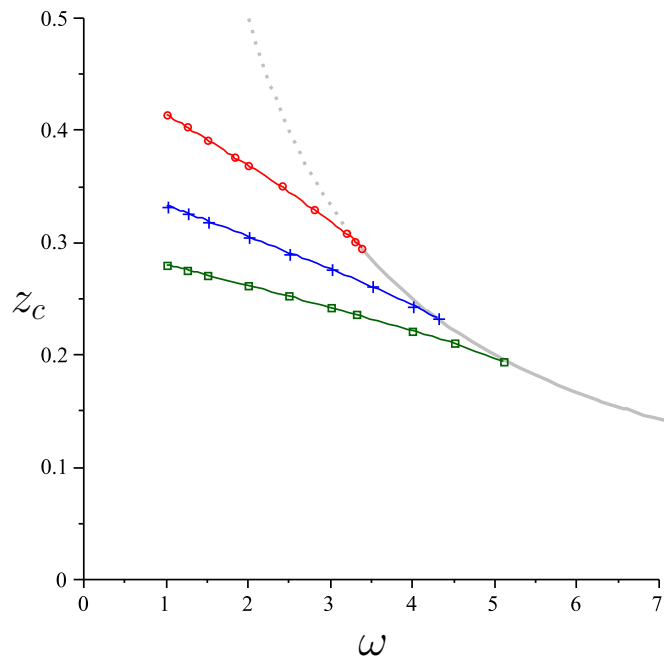}
  \caption{  
The radius of convergence of the generating function as
a function of $\omega$  when the force
is applied in the $x$-direction.  The rectangular hyperbola is independent of
the value of $h$.  The three curves with points marked correspond to
$h=1$ (top curve), $h=1.5$ and $h=2.0$.
}  
\label{pd-horiz}  
\end{figure}

To evaluate the continued fraction $\cal{C}$ efficiently we note that if
${\cal{C}}_m$ is the truncation of $\cal{C}$ at order $m$ then
${\cal{C}}_m$ can be written as a rational function whose numerator and 
denominator depend on $m$ and are determined by recurrences.  This gives a 
convenient way to evaluate $\cal{C}$ to the required accuracy.
In figure \ref{pd-horiz} we show the 
phase boundaries in the $(\omega,z)$-plane for several values of $h$.


\section{Extension under a fixed horizontal force}
\setcounter{equation}{0}
\label{exten-force-horiz}

We shall now consider the average extension of the polymer as a
function of the applied force at different temperatures. 

\begin{figure}[ht!]
  \psfrag{bbb}{\begin{large}$\beta^{-1}$\end{large}}
  \psfrag{fxt}{\begin{large}$f_x^t$\end{large}}
  \centering
  \includegraphics[width=9cm]{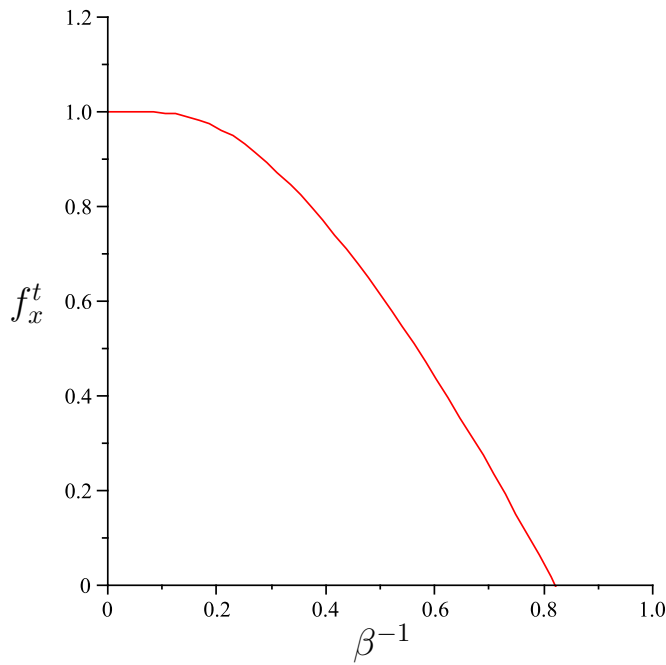}  
  \caption{  
The temperature dependence of the critical force for
horizontal pulling in the $n\to \infty$ limit.  When the force is
less than the critical force the walk is compact and when it is greater
than the critical force it is expanded.
}  
\label{ft-horiz}  
\end{figure}

From section~\ref{x-pull} we know that $\kappa(f_x,0,\beta)$ is a
continuous function of $\beta$ and $f_x$. Moreover for any fixed $f_x$
the free energy $\kappa(f_x,0,\beta)$ is analytic for $0 < \beta
<\infty$ except at at most one point which we label $\beta^t(f_x)$ when
it exists. This transition point is defined (see Owczarek and
Prellberg 2007) via
\begin{equation}
\omega_t = \left(\frac{\omega_t + h}{\omega_t-h}\right)^2
\label{crit-omega-horiz}
\end{equation}
recalling that $h=e^{\beta f_x}$.
Similarly  the free energy
$\kappa(f_x,0,\beta)$ is an  analytic function of $f_x$ except at the
solutions of $\beta^t(f_x,0) = \beta$. Consideration of the equation
(\ref{crit-omega-horiz}) implies there is a single solution.
Let the solution for this `critical force' be labelled $f^t_x$ and so by
(\ref{crit-omega-horiz}) it is given by the function
\begin{equation}
f_x^t =\frac{1}{\beta} \log\left( \frac{e^{\beta/2} - 1}{e^{-\beta/2} + e^{-\beta}}\right) \;,
\end{equation}
which is the result (11) of Rosa {\it et al.\ }(2003).  
In figure~\ref{ft-horiz} we plot the critical force $f_x^t$ against~$\beta^{-1}$. We note that for
$\beta \leq \beta^t(0,0)\approx 1.218$ (equivalently $T^t \approx
0.8205$) there is no positive critical force, and that
for $\beta > \beta^t(0,0)$ there is a single critical force $f_x^t$.

At fixed force and below the critical temperature (which depends on the force), we have
$z_c(h,1,\omega) = 1/\omega$ and so
\begin{equation}
\kappa(f_x,0,\beta) = \beta
 \quad \mbox{ for }   \beta \geq \beta^t(f_x,0).
\end{equation}
For $0< \beta \leq \beta^t(f_x,0)$ we know that $z_c(h,1,\omega)$ is a strictly
decreasing function of $\beta$, and so $\kappa(f_x,0,\beta)$ is a
strictly increasing function of $\beta$. Near the transition we know
from Owczarek and Prellberg (2007) (see the discussion after equation
(3.9)) that
\begin{equation}
\kappa(f_x,0,\beta) - \beta \sim C (\beta^t - \beta)^{3/2}  \quad 
\mbox{  as  } \beta \rightarrow (\beta^t)^-
\end{equation}
and similarly
\begin{equation}
\kappa(f_x,0,\beta) - \beta \sim D (f_x - f_x^t)^{3/2}  \quad 
\mbox{  as  } f_x \rightarrow (f_x^t)^+
\label{kappa-force-crit}
\end{equation}
for some constants $C$ and $D$.

For $0< \beta < \beta^t(f_x,0)$, the singularity in $z$ closest to the
origin in $\hat{G}(z;h,1,\omega)$ is a simple pole. From this we deduce that
\begin{equation}
Z_n(f_x,0,\beta) = e^{\kappa(f_x,0,\beta) n + O(1)} \quad \mbox{ for }   0< \beta < \beta^t(f_x,0)\;.
\label{part-scal-high-temp}
\end{equation}

In the semi-continuous model (Owczarek {\it et al.\ }(1993)) the following
was derived for the scaling of the partition function at low
temperatures
\begin{equation}
Z_n(f_x,0,\beta) = e^{\beta n + \kappa_s(f_x,0,\beta) n^{1/2}
  + O(1)} \quad \mbox{ for }   \beta^t(f_x,0) < \beta \;,
\label{part-scal-low-temp}
\end{equation}
where $\kappa_s(f_x,0,\beta)$ is a negative non-constant analytic
function of $\beta$ and $f_x$, and also we have that $\lim_{\beta
  \rightarrow \beta^t}\kappa_s(f_x,0,\beta^t)=0$. Both the discrete
model (given explicitly by Owczarek and Prellberg (2007)) and the
semi-continuous model (given by Owczarek {\it et al.\ }(1993)) have
generating functions that have uniform asymptotics with the same
algebraic structure, given by a ratio of Airy functions. In particular
the results in Owczarek {\it et al.\ }(1993) depend on the scaling
of the locations of the poles in the generating function which has the
same scaling in both cases. Without making the explicit calculation one can
surmise that equation (\ref{part-scal-low-temp}) also holds for the
discrete model.

For completeness we note that at $\beta=\beta^t(f_x,0)$ there is an algebraic
singularity in the generating function (see Owczarek and Prellberg
(2007)). Assuming the conditions of Darboux's theorem hold gives us
\begin{equation}
Z_n(f_x,0,\beta^t(f_x,0)) = e^{\beta^t n -\frac{2}{3} \log(n) + O(1)}\;.
\label{part-scal-critical}
\end{equation}

One could calculate the average extension in the generalised ensemble simply by
differentiating the generating function with respect to the force.  In the
discrete model this gives a complicated expression in terms of
$q$-Bessel functions and its derivatives. However as the equivalence
of the thermodynamic limits of the canonical ensemble and generalised
ensemble is limited to $0< \beta \leq \beta^t$ it is simpler to use
the results above to deduce the behaviour of the average extension in
the canonical ensemble directly. In the canonical ensemble the 
average extension
$\langle s_x\rangle_n(f_x,f_y,\beta)$ is defined by
\begin{equation}
\langle s_x\rangle_n(f_x,f_y,\beta) 
=  \frac{1}{\beta}
\frac{\partial \log Z_n(f_x,f_y,\beta) }{\partial f_x}\;.
\end{equation}
Assuming the conditions for differentiating the asymptotic expansion
term-by-term hold the result (\ref{part-scal-high-temp}) implies that
\begin{equation}
\langle s_x\rangle_n(f_x,0,\beta) 
= \frac{1}{\beta}
\frac{\partial \kappa(f_x,0,\beta)}{\partial f_x} n + O (1) \quad
\mbox{ for }   0< \beta < \beta^t\;,
\end{equation}
so that the thermodynamic limit extension per unit length
\begin{equation}
S_x(f_x,f_y,\beta) = \lim_{n \rightarrow\infty} \frac{\langle s_x\rangle_n(f_x,f_y,\beta)}{n}
\end{equation}
is 
\begin{equation}
S_x(f_x,0,\beta) = 
\frac{1}{\beta} 
\frac{\partial \kappa(f_x,0,\beta)}{\partial f_x} > 0 \mbox{ for }   0< \beta < \beta^t\;.
\end{equation}
Moreover the scaling near $\beta^t$ in equation (\ref{kappa-force-crit}) as  
given by Owczarek and Prellberg (2007)
implies that
\begin{equation}
S_x(f_x,0,\beta)  \rightarrow 0^+ \quad \mbox{ as } \beta \rightarrow (\beta^t)^-\;.
\end{equation}
One can calculate this quantity from the generalised ensemble in the
semi-continuous model (see equation (3.46) in Owczarek {\it et al}
1993) and analyse it in a more straightforward manner than in the
discrete case (since it involve Bessel functions rather than
$q$-Bessel functions), and it shows the same behaviour.

On the other hand the result (\ref{part-scal-low-temp}) (once again
assuming differentiability of the asymptotic expansion) implies that 
\begin{equation}
\langle s_x\rangle_n(f_x,0,\beta) 
= \frac{1}{\beta}
\frac{\partial \kappa_s(f_x,0,\beta)}{\partial f_x} n^{1/2} + O (1) \quad
\mbox{ for }   \beta^t < \beta 
\end{equation}
and so that
\begin{equation}
S_x(f_x,0,\beta) = 0 \quad
\mbox{ for }   \beta^t < \beta \;.
\end{equation}

We have calculated the stress-strain curves by computing the free energy 
from the boundary of convergence (section 4) and then numerically 
differentiating.  Using these results and the above arguments we
can make the following comments.
In all cases the function of
the average extension per unit length $S_x(f_x,0,\beta)$ is a continuous
function of $f_x$ for $f_x\geq 0$.

At high temperatures where $\beta < \beta^t(0,0)$ there is no critical
force and the average extension per unit length $S_x(0,0,\beta)>0$ at zero
force. At such fixed $\beta$ the function $S_x(f_x,0,\beta)$ is an
analytic function of $f_x$ for $f_x\geq 0$ and is strictly increasing
with increasing force. There is a unit horizontal asymptote,  approached
from below for large forces. There is a finite
nonzero slope $\frac{\partial S_x(f_x,0,\beta)}{\partial f_x}$ at $f_x=0$. 
The qualitative behaviour is the same as that shown in figure~\ref{fig ss w1}
for the the $\omega = 1$ case.

 \begin{figure}[ht!]
 \psfrag{ss}[b][][1][90]{\begin{large}$\ds \lim_{n \to \infty} \frac{\langle s_x
  \rangle_n}{n}$\end{large}}
 \psfrag{bb}[]{\begin{large}$\beta f_x$\end{large}}
  \begin{center}
   \includegraphics[height=8cm]{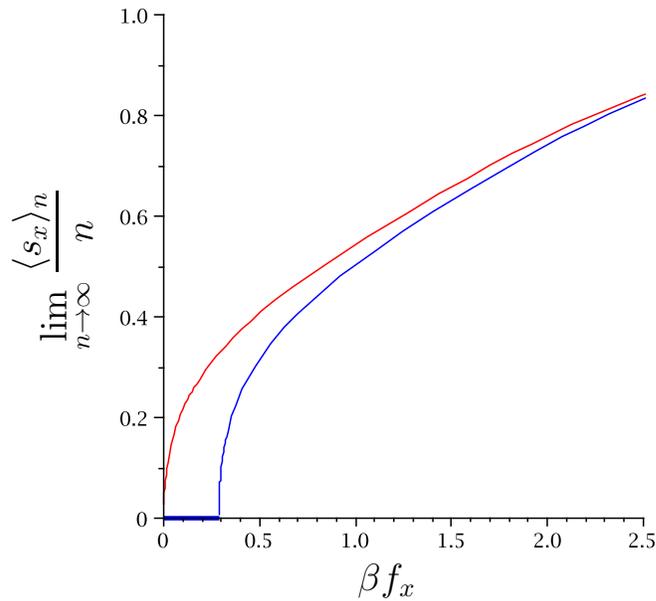}
  \end{center}
  \caption{The dependence of $\ds \lim_{n\to\infty} \langle s_x \rangle_n / n$, the limiting average
  horizontal span per unit length, on $\beta f_x$ for
  different temperatures: at $\beta=\beta_c$ (upper curve) and at
  $\beta=\log 4>\beta_c$ (lower curve). Note that above the critical temperature
  the behaviour is qualitatively the same as that shown in figure~\ref{fig
  ss w1}.}
  \label{fig svsf both}
 \end{figure}

At the critical temperature for zero force $\beta = \beta^t(0,0)$ the
average extension per unit length $S_x(0,0,\beta)=0$ at zero force. Again
the function $S_x(f_x,0,\beta^t(0,0))$ is an analytic function of $f_x$ for
$f_x>0$, is strictly increasing with increasing force, and has a
unit horizontal asymptote, approached from below for large forces.
Now however the slope 
$\frac{\partial S_x(f_x,0,\beta^t(0,0))}{\partial f_x}$ diverges when
$f_x \rightarrow 0^+$ as $f_x^{-1/2}$. This is shown in figure~\ref{fig svsf
both}.

At low temperatures where $\beta>\beta^t(0,0)$ (so there is a
critical force $f_x^t >0$) there are two regimes. For $0\leq f_x\leq
f_x^t$ the average extension per unit length $S_x(f_x,0,\beta)=0$
regardless of the force. However for $f_x^t< f_x$ the function
$S_x(f_x,0,\beta)$ is an analytic function of $f_x$, is strictly
increasing with increasing force, and has a unit horizontal asymptote,
approached from below for large forces. Again the slope
$\frac{\partial S_x(f_x,0,\beta)}{\partial f_x}$ diverges when $f_x \rightarrow (f_x^t)^+$
as $(f_x-f_x^t)^{-1/2}$. This is illustrated in figure~\ref{fig svsf
both}.


\section{The full model: Pulling in both directions }
\setcounter{equation}{0}
\label{both-pull}

\subsection{No self-interactions ($\omega=1$)}
As in subsection~\ref{x-pull-no-int} which describes the solution when there is no vertical pulling (and no self-interactions) one can write down a simple functional equation for the generating function $G(x,y_+,y_-,1)$ as
\begin{equation}
G(x,y_+,y_-,1) =  x (1 + \frac{y_+}{1-y_+} + \frac{y_-}{1-y_-}) (1+G(x,y_+,y_-,1))
\end{equation}
so that
\begin{equation}
1 + G(x,y_+,y_-,1) = \frac 1 {1 - x \frac{1-y_+y_-}{(1-y_+)(1-y_-)}}\;,
\end{equation}
and setting $x=hz$, $y_+=zv$ and $y_-=z/v$ we get
\begin{equation}
1 + \hat G(z;h,v,1) = \frac 1 {1 - hz \left[\frac{1-z^2}{1-(v+v^{-1})z+z^2}\right]}\;.
\end{equation}
So the answer is a simple rational function, and the critical $z$ (that is $z_c(h,v,1)$) is the
root of a cubic. For $h=v=1$ this simplifies as expected.

\begin{figure}[ht!]  
  \psfrag{bfy}{\begin{large}$\beta f_y$\end{large}}
    \psfrag{lsn}{\begin{large}$\ds \lim_{z\to z_c}
\frac{\langle s_y\rangle}{\langle n \rangle}$\end{large}}
  \centering
  \includegraphics[width=9cm]{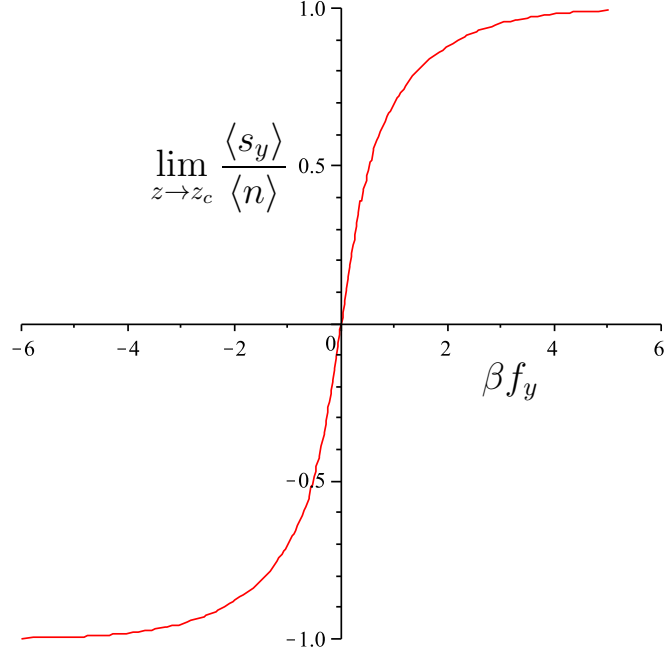}
  \caption{A plot of $\ds \lim_{z \to z_c} \langle
s_y\rangle /\langle n \rangle$ when $\omega=1$ against $\beta f_y$.}
\label{fig ss w1 vert}  
\end{figure}

If we set $f_x=0$ then
\begin{equation}
1 + \hat G(z;1,v,1) = \frac 1 {1 - z \left[\frac{1-z^2}{1-(v+v^{-1})z+z^2}\right]}\;,
\end{equation}
and from this we can readily calculate the ratio 
$\langle s_y \rangle /\langle n\rangle$ and take the 
thermodynamic limit by letting $z\to z_c(1,v,1)$. In figure~\ref{fig ss w1 vert} is a plot of
$\lim_{z \to z_c} \langle s_y\rangle /\langle n \rangle$ against $\beta f_y$.

\subsection{Solution of the full model}
In the appendix of Owczarek {\it et al.\ }(1993) it was shown that if one adds
fugacity variables $y_+$ for steps in the positive vertical direction
and $y_-$ for steps in the negative vertical direction then a
generalisation of the method for solving the standard problem yields
the generating function $G(x,\yp,\ym,\w)$ as described in
section~\ref{model}. Using $\qp=\yp\omega$ and $\qm=\ym\omega$ it is given by
\begin{equation}
1+G(x,\yp,\ym,\w)=\frac{\left(1-\w\right)}{\left[2\bar{\cH}(x,\yp,\ym,\w)-
(1+\w+ (1-\w)x)\right]}\;,
\label{solution-vert}
\end{equation}
where
\begin{eqnarray}
\label{solution2}
\bar{\cH}(x,\yp,\ym,\w)
&=&\frac{(\Ap_0+\Bp_0)(\Ap_1-\Bp_1)-(\Am_0+\Bm_0)(\Am_1-\Bm_1)}{(\Ap_0+\Bp_0)(\Ap_0-\Bp_0)-(\Am_0+\Bm_0)(\Am_0-\Bm_0)}\;,
\end{eqnarray}
with
\begin{eqnarray}
\Apm_r&=&\sum_{m=0}^\infty\frac{x^{2m}(\w-1)^{2m}(\qp\qm)^{m(m+r)}\qpm^m}
{\prod_{k=1}^mP[(\qp\qm)^{k-1}\qpm]P[(\qp\qm)^k]}\\
\Bpm_r&=&\sum_{m=0}^\infty\frac{x^{2m+1}(\w-1)^{2m+1}(\qp\qm)^{m(m+r)}\qpm^{r+m+1}}
{P[(\qp\qm)^m\qpm]\prod_{k=1}^mP[(\qp\qm)^{k-1}\qpm]P[(\qp\qm)^k]}\;,
\end{eqnarray}
and 
\begin{equation}
P[\l]=(\l-1)(\l-\w).
\end{equation}
Note that $\qp\qm=q^2$, and given that all the parameters are positive
we have $q=\sqrt{\qp\qm}$.
This solution is then a clear generalisation of the form for the horizontal pulling model
 where
 \begin{equation} 
\label{solution}
1+G(x,y,y,\w)=\frac{1-\w}{\left[2\cH(y,y\w,xy(\w-1))- (1+\w+ (1-\w)x)\right]}.
\end{equation}
where $\cH(y,q,t)= H(y,q,qt)/ H(y,q,t)$ is given in terms of
$H(y,q,t)$ defined by equation~(\ref{Hfunction}). 

However the problem of generalising the analysis of Prellberg (1995)
is a daunting one. On the other hand we can still make some deductions
about the behaviour of the generating function. Since the partition
function is positive we know that $\hat{G}(z;h,v,\omega)$ is a
strictly increasing function of $z$ for fixed $h,v,\omega > 0 $. 

Let us consider $\omega\geq 1$. The functions $\Apm_r$ and $\Bpm_r$
converge whenever $q < 1$ and moreover there are singularities when
$|q|=1$. The solution for the generating function $G(x,\yp,\ym,\w)$ is
given in the next section and one finds that $G(x,\yp,\ym,\w)$ on this
surface is finite when $\omega > \omega_t$, where $\omega_t$ is given
by the solution of equation (\ref{location-crit-point}). The only
singularities that can occur in the generating function for $q < 1$ 
are poles 
occurring when the denominator of (\ref{solution-vert}) is zero. The nature of
these poles is unclear but we conjecture that they are simple poles and
that they only exist when $\omega < \omega_t$.

\subsection{Solving the full model on the special surface $z=1/\omega$}

While expression (\ref{solution2}) is rather unwieldy, it is possible to
find a sub-manifold of parameters for which the solution can be made more
explicit. If we restrict to 
\begin{equation}
q_+q_-= q^2 = 1 \quad \mbox{ that is } q=\omega z= 1\;,
\end{equation}
then  (\ref{solution2}) has a singular limit. This is analogous to
the fact that in the case of (\ref{solution}) one obtains an algebraic 
function for $q=1$. 

Hence we have $q_+=v$ and $q_-=1/v$. Also, we shall retain the variable $x$ for convenience
during the calculations before substituting $x=h/\omega$ as required in
the final expressions.

Let  $\gpm_r=\ypm^{-r}\Gpm_r$ where $\Gpm_r$ is the generating 
function for walks with $r$ steps in the positive and  
negative directions respectively, and we have in 
analogy with the derivation of the full  problem
\begin{eqnarray} 
\label{constant_coefficients}
\gpm_{r+4}-(\w+1)(\qpm+1)\gpm_{r+3}+(\w(1+\qpm^2 )& + &(\w+1)^2\qpm)\gpm_{r+2}\\
-\w\qpm(\w+1)(\qpm+1)\gpm_{r+1}+\w^2\qpm^2\gpm_r
& = &\qpm x^2(\w-1)^2\gpm_{r+2}\;. \nn
\end{eqnarray}
The characteristic polynomial is
\begin{equation}
P_{\pm}(\lambda)=(\lambda-1)(\lambda-\w)(\lambda-\qpm)(\lambda-\qpm\w)-\qpm x^2(\w-1)^2\lambda^2\;.
\label{charpoly}
\end{equation}
Despite being a polynomial of degree four, $P_{\pm}(\lambda)$ has sufficient 
symmetry to allow for simple explicit solutions. The key observation is that
if $\lambda$ is a root of $P_{\pm}(\lambda)$, then so is $\qpm\w/\lambda$.
If we define
\begin{equation}
\label{muoflambda}
\mu_{\pm}=\lambda+\frac{\w\qpm}\lambda\;,
\end{equation}
then $\mu_{\pm}$ satisfies
\begin{equation}
\label{mueqn}
\mu_{\pm}^2-(\w+1)(\qpm+1)\mu_{\pm}+(\w+\qpm)(1+\w\qpm)-x^2(\w-1)^2 =0\;.
\end{equation}
We can therefore obtain all four roots of $P_{\pm}(\lambda)$ by 
solving the quadratic equation (\ref{mueqn}) for $\mu_{\pm}$, followed
by solving the quadratic equation (\ref{muoflambda}) for $\lambda$.

The general solution for (\ref{constant_coefficients}) is a linear combination
\begin{equation}
\gpm_r=\sum_{i=1}^4A_i\lambda_{\pm,i}^r\qquad\mbox{respectively}\qquad
\Gpm_r=\ypm^r\sum_{i=1}^4A_i\lambda_i^r\;.
\end{equation}

After some algebra the generating function $ 1 +
\hat{G}(1/\omega;h,v,\omega)= G(h/\omega,v/\omega,1/v\omega,\omega)$ can be found to be
\begin{eqnarray}
1 + \hat{G}(1/\omega;h,v,\omega)=  
 \frac{(t_1t_2\omega-1)(t_1v-1)(t_2v-1)(\omega-1)\omega^2}{D(t_1,t_2)}\;,
\label{vertical-special}
\end{eqnarray}
where
\begin{eqnarray}
D(t_1,t_2)&=& t_1t_2(\omega-1)^2h^2+(\omega-1)(t_1t_2v-1)(t_1t_2\omega v-1)h\omega\nonumber\\
&-& (t_1-1)(t_1v-1)(t_2-1)(t_2v-1)\omega^3
\end{eqnarray}
with $t_1$ and $t_2$ the  two distinct roots of $L(t)=0$ with
\begin{equation}
L(t)= v h^2 (\omega-1)^2 t^2 - (t-1)(\omega t -1) (v t -1)
(\omega v t-1)\omega^2\;.
\end{equation}
We note that $L(t)$ is a quartic polynomial in $t$.
If we let 
\begin{equation}
s = \frac{1}{t} + v\omega t
\label{s-of-t}
\end{equation} 
then $s$ satisfies
\begin{equation}
s^2 -(\omega+1)(v+1) s + (\omega+v)(1+\omega v) = v h^2 \frac{(\omega-1)^2}{\omega^2}\;,
\end{equation}
and the two solutions of this equation are
\begin{equation}
s_\pm = \frac{(\omega+1)(v+1)\pm\sqrt{(\omega-1)^2\left(v^2-2v+1+4vh^2/\omega^2\right)}}{2}\;.
\end{equation}
Now only one of the solutions of (\ref{s-of-t}) is applicable, namely
\begin{equation}
t = \frac{s-\sqrt{s^2-4v\omega}}{2v\omega}
\end{equation}
which gives us that
\begin{equation}
t_1= \frac{s_+-\sqrt{s_+^2-4v\omega}}{2v\omega}
\end{equation}
and
\begin{equation}
t_2= \frac{s_- -\sqrt{s_-^2-4v\omega}}{2v\omega}\;.
\end{equation}
Importantly, on substitution into the generating function the
denominator can be found to be a polynomial with a factor
\begin{equation}
\left(v\omega^2+2h\omega v+v
  h^2-2v h^2\omega+\omega^4v-2h\omega^3v+v h^2\omega^2-\omega^3-v^2\omega^3\right) 
\end{equation}
 and no other relevant factors. Hence the generating function is 
 singular on the curve $q=1$ when this factor is zero. It can be seen
 that the algebraic singularity (a square root) occurs in the
 generating function at the same place by considering the discriminant of (\ref{charpoly}).
It can therefore be deduced that the exponent $\gamma_u=1/2$
describing the singularity in the generating function approaching the
transition point tangentially (as described in Owczarek {\it et al}
1993). The exponent is independent of the value of $h$ and $v$ and
so of whether there is a horizontal and/or vertical pulling force.
 
\begin{figure}[ht!]
  \psfrag{bb}{\begin{large}$\beta^{-1}$\end{large}}
  \psfrag{fyt}{\begin{large}$f_y^t$\end{large}}
  \centering
  \includegraphics[width=8cm]{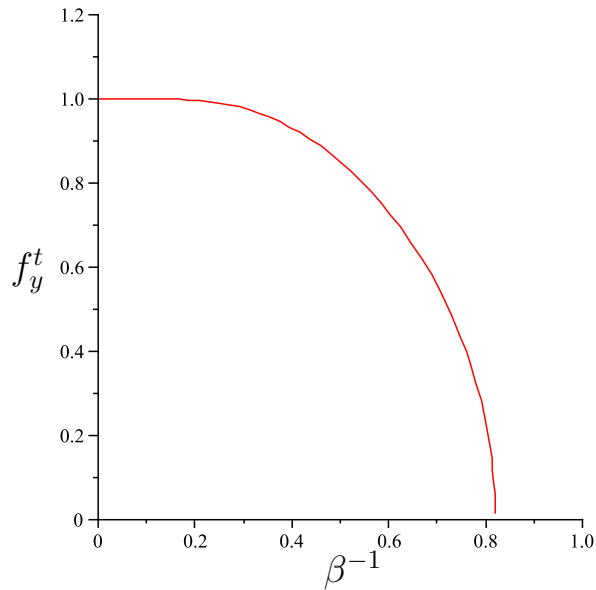}  
  \caption{  
  The temperature dependence of the critical force for
  vertical pulling in the $n\to \infty$ limit.}
  \label{ft-vert}
\end{figure}

We therefore have the location of the critical point as 
\begin{equation}
\omega^2(1+\omega^2) +2\omega(1-\omega^2)h + (1-\omega)^2h^2 =
\omega^3 (v +1/v)\;.
\label{location-crit-point}
\end{equation}
One can rewrite this as
\begin{eqnarray}
\cosh(\beta f_y/2)  = &\exp(-\beta/2) \cosh(\beta)\nonumber\\
                   &- \exp(-\beta) \sinh(\beta/2) (\exp(\beta f_x)-1)\;.
\label{full-force-temp}
\end{eqnarray}
When there is no horizontal force, that is $h=1$, we therefore find
 the critical point to be when
\begin{equation}
(\omega^2+1)^2 = (v+2+\frac{1}{v})\omega^3\;,
\end{equation}
which reduces to the known result of Binder {\it et al.\ }(1990) when
$v=1$.  From this equation the critical force temperature plot can
be found as
\begin{equation}
f^t_y = \beta^{-1} \cosh^{-1} \left( 2e^{-\beta }\cosh^2(\beta) -1  \right).
\end{equation}
In figure~\ref{ft-vert} we plot the critical force against temperature for
vertical pulling. Note that as the temperature approaches the critical
value for no pulling force the slope of this curve diverges in
contrast to the analogous horizontal pulling curve.

Note that when $v=1$ in (\ref{location-crit-point}) we find
equation (\ref{crit-omega-horiz}), that was derived in a previous section.



\section{Functional equation method}
\setcounter{equation}{0}
\label{func-eqns-vert}

\newcommand{\cP}{\mathcal{P}}
\newcommand{\ntlm}{second column vertical}

\newcommand{\cn}{\omega}        
\newcommand{\vs}{y}                             
\newcommand{\ls}{p}                             
\newcommand{\ts}{z}                             
\newcommand{\hs}{x}                             

\subsection{Horizontal pulling}
\label{sec horiz pull}

For the case of horizontal pulling we can construct a recursive 
functional equation for the partial generating function, \emph{cf}
section~\ref{model}. 
Let us define 
\begin{equation}
g_r(\hs,\vs, \cn)
 =  \left\{\begin{array}{cc}
2G_r(x,y,y,\omega) & \textrm{if $r\ge 1$}\\
 G_0(x,y,y,\omega) & \textrm{if $r=0$}
\end{array}\right.
\label{eq_dgdg1}
\end{equation}
so that
\begin{eqnarray}
F(\ls)&=\sum_{r\ge 0} g_r \ls^r \;.
\label{eq_tempfunc}
\end{eqnarray}
\begin{figure}[htbp]
        \begin{center}
   \includegraphics[width=14cm]{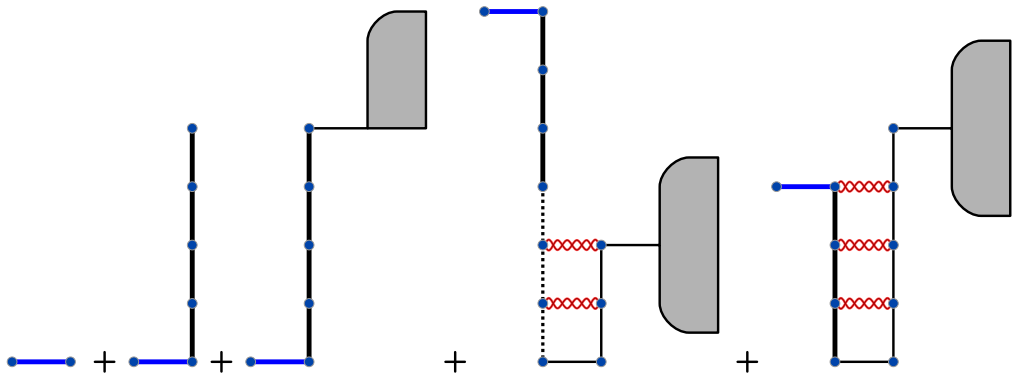}
        \end{center}
        \caption{Schematic representation of the terms of the functional
equation (\ref{eq_kerhsor}) arising from the five cases described in
section~\ref{sec horiz pull}.}
        \label{fig_figures_kernalHorizontal}
\end{figure}
%
The functional equation is constructed by considering what happens
when an extra column is added to the left of a walk. Let $\cP_r$ be
the set of all paths with at least one horizontal step followed by
$r\ge 0$ first column vertical steps. Let $r'$ be the number of \ntlm\ 
steps. We can partition $\cP_r$ into the following five disjoint
subsets.

\paragraph{Case I:} The walk has only one step. This is generated by $x$.

\paragraph{Case II:} These are all walks with only one horizontal step
and at least one first column vertical step. These are generated by
$2\hs {  \vs  \ls}/({1-\vs \ls})$, the factor of two giving all
upwards and all downwards sequences of steps, both generated by $ {
  \vs  \ls}/({1-\vs \ls})$. 

\paragraph{Case III:} These are all paths with at least two horizontal
steps and  the  first column vertical steps  (if any) are in the
\emph{same} direction to the \ntlm\ steps (if any). These are
generated by $\hs F(1)/({1-\vs \ls})$. There is no factor of two for
the following reason.  The product $\hs F(1)/({1-\vs \ls})$
corresponds to concatenating a sequence of vertical steps, $V\to
1/({1-\vs \ls})$ to and arbitrary path  $ F(1)$. The generating
variable $p$ is set to one in $F(p)$ since the leftmost vertical steps
arise from the $V$ sequence.  If the first vertical sequence of steps
of $ F(1)$ are upward, then the $V$ steps are interpreted as also
being upward (and hence no contacts are created by the concatenation,
thus no $\cn$ factor). If the first vertical sequence of steps of $
F(1)$ are downward, then the $V$ steps are interpreted as also being
downward (and again, no contacts are created by the concatenation). 

\paragraph{Case IV:} These are all paths with at least two horizontal steps and $0<r\le r'$ and with
the further condition that the vertical steps in the first column are
in the \emph{opposite} direction to those in the second column. These
are generated by $\hs \vs \ls F(\cn \vs \ls)/({1- \vs \ls})$. The
argument for this form is similar to that of Case III, except now
contacts are created by the concatenation. The contacts are accounted
for as follows. All new contacts occur in the overlap between the
first and second columns. Since the $p$ in $F(p)$ tracks the number of
vertical steps in the second column, the new contacts can be accounted
for by replacing $p$ by $\cn\ls$. The factor of $yp$ in $F(\cn\vs
\ls)$ is interpreted as giving rise to first $r'$ vertical steps in
the \emph{first} column (these can be thought of as `virtual steps')
and the remaining $r'-r$ vertical steps are generated by $1/({1-\vs
  \ls})$ and are interpreted as being concatenated on to the virtual
steps, as illustrated by the fourth term in
figure~\ref{fig_figures_kernalHorizontal}.

\paragraph{Case V:} These are all paths with at least two horizontal steps and
$0<r'\le r $ and with the further condition that vertical steps in the first column are in the
\emph{opposite} direction to those in the second column. These are generated by $ \hs 
\cn \vs \ls \left[F(1)-F(\cn \vs \ls)\right]/({1-\cn \vs  \ls  })$ which can be shown using
inclusion-exclusion. The term $\hs  \cn\vs \ls /({1-\cn \vs \ls  })$ counts an infinite sequence of
vertical bonds and contacts. So the generating function, $T_1= \hs  \cn\vs \ls F(1)/({1-\cn \vs 
\ls })$, counts all the required configurations, but also those with $r'>r$. The
contribution of these over-counted configurations is then given by $T_2= \hs  \cn\vs \ls F(\cn\vs
\ls)/({1-\cn \vs \ls })$. Hence the contribution of this case is $T_1-T_2$. See figure~\ref{fig
feqn caseV}.
\begin{figure}[ht!]
  \begin{center}
  \includegraphics[width=12cm]{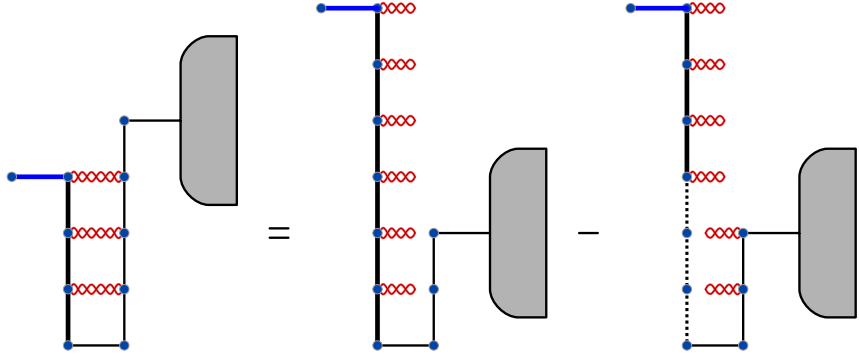}
  \end{center}
  \caption{Schematic representation of contribution of Case V to the
functional equation.}
  \label{fig feqn caseV}
\end{figure}

Combining all five cases together,  as illustrated in
figure~\ref{fig_figures_kernalHorizontal},   generates all the walks. Thus we
get,
\begin{eqnarray}
        F(\ls)= \hs &+& 2\hs\frac{  \vs  \ls}{1-\vs \ls}+\hs\frac{1 }{1-\vs \ls}F(1)+\hs\frac{ \vs \ls}{1- \vs \ls}F(\cn\vs \ls)\nonumber\\
       &+& \hs \frac{\cn\vs \ls}{1-\cn\vs  \ls }\biggl[F(1)-F(\cn\vs \ls)\biggr].
        \label{eq_kerhsor}
\end{eqnarray}
This functional equation can be solved for $F(\ls)$ by the method of
iteration (Bousquet-M\'elou~1996).

The case $ \cn\vs  =1$ is much simpler to solve. Letting $ \cn\vs  =1$ gives,
\begin{eqnarray}
        F(\ls)= \hs &+& 2\hs\frac{  \vs  \ls}{1-\vs \ls}+\hs\frac{1 }{1-\vs \ls}F(1)+\hs\frac{ \vs \ls}{1- \vs \ls}F( \ls)\nonumber\\
        &+&\hs \frac{ \ls}{1- \ls  }\biggl[F(1)-F ( \ls)\biggr].
        \label{eq_kerhsorsimp}
\end{eqnarray}
Note, to find $F(1)$ from equation (\ref{eq_kerhsorsimp}) we cannot put $p=1$ because of the denominator $1-p$. To avoid this problem, first   collect coefficients of $F(p)$ to give,
\begin{equation}
 K(\ls)  F(\ls)= \hs (1-\ls)(1+\vs\ls) + \hs  (1-\vs\ls^2)F(1)
\label{eq_q1l1}
\end{equation}
where the ``kernel'', $K(p)$, is given by
\begin{equation}
K(\ls)= y \ls^2-(1-\hs+\vs+\hs\vs)\,\ls+1.
\label{eq_q1l2}
\end{equation}
If we can put $K(p)=0$ in (\ref{eq_q1l1}) then, since the lefthand side is zero,   we can solve for $F(1)$. Note, the condition $K(p)=0$ implicitly constrains $p$, which, since $K(p)$ is quadratic in $p$, implies that we get  two possible functions $\ls_\pm(\hs,\vs)$.   However some care must be taken as this way of finding $F(p)$  assumes $F(p)K(p)=0$ which is only the case if $\lim_{p\to p_\pm}F(p)K(p)=0$. Thus, we first assume   this is the case for one of the solutions, say $p_+$  and then verify this assumption once $F(p) $ has been explicitly computed. 
As is readily shown, each of the two assumptions $\lim_{p\to p_+}F(p)K(p)=0$ and $\lim_{p\to p_-}F(p)K(p)=0$ give rise to \emph{different} functions, denoted $F_\pm(p)$. In fact, $F(p)$ is an algebraic function and each of the two functions $F_\pm(p)$ are the two branches of $F(p)$. The limit $\lim_{p\to p_+}F(p)K(p)$ only vanishes if the correct branch is combined with the correct limit (the other branch has a pole at $p_+$ and hence the limit is the non-zero residue of $F(p)$ at $p_+$).

Thus, assuming $\lim_{p\to p_+}F_+(p)K(p)=0$ and taking the limit as $\ls\to\ls_+$ on both sides of (\ref{eq_q1l1}) gives   $F_+(1)$. Thus we obtain the branch $F_+(1)$ on the line $\cn\vs  =1$ as
\begin{equation}
        F_+(1)=-\frac{(1-\ls_+)(1+\ls_+\vs)}{1-\ls^2_+\vs}
\end{equation} 
        or, explicitly
\begin{equation} 
        1+F_+(1)=\sqrt{\frac{\vs-1}{\hs^2(\vs-1)+2\hs(\vs+1)+\vs-1}}.
\end{equation} 
Using this solution, one then verifies the assumption that $\lim_{p\to p_+}F_+(p)K(p)=0$. This solution has a square root singularity  at
\begin{equation}
          \hs^2(\vs-1) +2 \hs(\vs + 1)+\vs-1=0,
\end{equation}
which gives the same critical value of $\hs$, $\hs_c$ as given by
equation (\ref{crit-omega-horiz}) using the transformations
$x=h/\omega$ and $y=1/\omega$. Thus we see that, using this method, $\hs_c$ arises via $p_+$ and hence from the kernel.

The functional equation (\ref{eq_kerhsor}) is also closely linked to the recurrence relation (\ref{sum-rec}). If we make the substitution (\ref{Fp}) into (\ref{eq_kerhsor}) and equate coefficients of $\ls$, then, for $r>0$ we get
\begin{equation}
        g_r=\hs\vs^{r+1}\left(2  +\sum_{k\ge 0}g_k+\sum_{k=0}^{r-1} \cn^k g_k+\cn^r\sum_{\k\ge0} g_k-\cn^r\sum_{k=0}^{r-1}g_k \right)
\end{equation}
which is readily put in the same form as (\ref{sum-rec}). 


\subsection{Vertical pulling}

\newcommand{\vv}{h}
\newcommand{\vd}{\bar{v}}                               
\newcommand{\vu}{v}
\newcommand{\gfu}{G^+}
\newcommand{\gfd}{G^-}
\newcommand{\fgfd}{\hat{F}^-}
\newcommand{\fgfu}{\hat{F}^+}
\newcommand{\fgfdb}{\overline{F}^-}
\newcommand{\fgfub}{\overline{F}^+}
\newcommand{\mattt}[1]{\mathbf{ #1}}
\newcommand{\vecg}{\mathbf{F}}

Let us define
\begin{equation}
\hat{F}^{+}(p) = \sum_{r=0}^{\infty} G_r^{+}(x,v y,y/v,\omega) p^r
\end{equation}
and
\begin{equation}
\hat{F}^{-}(p) = \sum_{r=1}^{\infty} G_r^{-}(x,y/\bar{v},y \bar{v},\omega) p^r\;.
\end{equation}

Using similar arguments to the derivation of equation (\ref{eq_kerhsor}) we obtain the following pair of coupled equations.
\begin{eqnarray}
        \fgfu(\ls)&=& \hs + 
        \hs \vu \frac{  \vs  \ls\vu }{1-\vs \ls\vu }+
        \hs\frac{1 }{1-\vs \ls}\fgfu(1)+
        \hs\frac{ \vs \ls\vu }{1- \vs \ls\vu }\fgfd(\cn\vs \ls\vu )\nonumber\\
& &\qquad +       \hs \frac{\cn\vs \ls\vu }{1-\vs  \ls \cn \vu }\biggl[\fgfd(1)-\fgfd(\cn\vs \ls\vu )\biggr]
\end{eqnarray}
and
\begin{eqnarray} 
        \fgfd(\ls)&=& 
        \hs  \frac{  \vs  \ls\vd }{1-\vs \ls\vd }+
        \hs\frac{1 }{1-\vs \ls\vd}\fgfd(1)+
        \hs\frac{ \vs \ls\vd }{1- \vs \ls\vd }\fgfu(\cn\vs \ls\vd )\nonumber\\
&&\qquad +       \hs \frac{\cn\vs \ls\vd }{1-\cn \vs  \ls \vd }\biggl[\fgfu(1)-\fgfu(\cn\vs \ls\vd )\biggr]\;.
\end{eqnarray}

These equations can again be solved by the method of iteration, resulting in complex $q$-series similar to (\ref{solution2}).  

There is also the simple case  occurring when $\cn\vs  =1$.  Using the notation
\[
L'(a)=\frac{1}{1-a},\qquad  L(a)=aL'(a)\qquad \mbox{and} \qquad\Delta L(a)=L(y a)-L(a)\;,
\]
we can write the two functional equations as
\begin{eqnarray}
        \fgfub(\ls)=1+L(\vs\ls \vu)&+&L'(\vs\ls\vu)      \fgfub(1)+L(\vs\ls\vu)  \fgfdb(\ls\vu)\nonumber\\
        &+&
        L(\ls\vu)\left[         \fgfdb(1)- \fgfdb(\ls\vu) \right]
\end{eqnarray} 
and
\begin{eqnarray} 
        \fgfdb(\ls)=L(\vs\ls \vd)&+&L'(\vs\ls\vd)        \fgfdb(1)+L(\vs\ls\vd)  \fgfub(\ls\vd)\nonumber\\
        &+&      L(\ls\vu)\left[         \fgfub(1)- \fgfub(\ls\vd) \right]\;,
\end{eqnarray}
where
\[
\hs\fgfub=      \left. \fgfu\right|_{\cn\vs  =1}\qquad\mbox{and}\qquad \hs\fgfdb=       \left. \fgfd\right|_{\cn\vs  =1}.
\]
Solving the two equations for $\fgfub(\ls)$ and $\fgfdb(\ls)$ gives the pair
\begin{eqnarray}
        K^+(\ls,\vu)\fgfub(p)=1&+&L(\vs\ls \vu)+\hs L(\vs\ls ) \Delta L(\ls\vu)\nonumber\\
        &+&\left[L'(\vs\ls \vu)+\hs  L(\ls )\Delta L(\ls\vu)   \right]\fgfub(1)\nonumber\\
         &+&\left[L(\ls \vu)+\hs  L'(\vs\ls )\Delta L(\ls\vu)   \right]\fgfdb(1)
\end{eqnarray}
and
\begin{eqnarray}
                K^-(\ls,\vu)\fgfdb(s)=L(\vs\ls \vd)&+\hs L(\vs\ls )\Delta L(\ls\vd)\left[1+L(\vs\ls ) \right]\nonumber\\
        &       +\left[L'(\vs\ls \vd)+\hs  L(\ls )\Delta L(\ls\vd)   \right]\fgfdb(1)\nonumber\\
                & +\left[L(\ls \vd)+\hs  L'(\vs\ls )\Delta L(\ls\vd)   \right]\fgfub(1) \;,
\end{eqnarray}
where the two kernels are given by
\begin{eqnarray}
        K^+(\ls,\vu)&=&(1-\ls)(1-\vs\ls)(1-\ls\vu)(1-\vs\ls\vu)-\hs^2\ls^2\vu(\vs-1)^2\\
        K^-(\ls,\vu)&=&K^+(p,\vd)\;.
\end{eqnarray}
Thus we choose $\ls$ such that 
\begin{equation}
         K^-(\ls(\vu),\vu)=K^+(\ls(\vd),\vd)=0,  
        \label{eq_kernvpz}
\end{equation}
which is the same equation as the characteristic equation (\ref{charpoly}) with $p\to \lambda$ and $y$ replaced by $1/\omega$. Thus we see that the kernel that arises from the functional equation approach corresponds to the characteristic equation (\ref{charpoly}) required to solve the Temperley recurrence relations (\ref{constant_coefficients}).

%
%


\section{Three-dimensional model }
\setcounter{equation}{0}
\label{3d-model}

The model can be generalized to a three-dimensional partially
directed walk model in which the walk is self-avoiding, can take 
steps in the $+x$ or $+y$ directions or in the $\pm z$ directions.
Again it will be convenient to require that the first step is in the 
$+x$ or $+y$ direction.  We shall only consider a force applied in the 
$(x,y)$-plane and as such we shall keep track of the span in the $x$ and $y$ 
directions as well as the number of contacts.

If we write $c_n$ for the number of these walks with $n$ steps it is
easy to see that
\begin{equation}
\lim_{n\to\infty} n^{-1} \log c_n = \log[(3+\sqrt{17})/2].
\end{equation}
Suppose that $c_n(s_x,s_y,m)$ is the number of walks with 
$n$ steps, $m$ contacts, and with spans $s_x$ and $s_y$ in the 
$x$ and $y$ directions.  The corresponding canonical partition 
function is
\begin{equation}
Z_n(h_x,h_y,\omega)=\sum_{s_x,s_y,m}c_n(s_x,s_y,m) h_x^{s_x}h_y^{s_y}\omega^m
\end{equation}
and we define the generating function
\begin{equation}
\hat{G}(z;h_x,h_y,\omega) = \sum_n Z_n(h_x,h_y,\omega)z^n.
\end{equation}

Concatenation arguments can be used, as in section~\ref{x-pull}, to establish the 
existence of the limiting free energy  
\begin{equation}
\kappa(h_x,h_y,\omega) = \lim_{n\to\infty} n^{-1} \log Z_n(h_x,h_y,\omega).
\end{equation}
Methods exactly analogous to those in section~\ref{x-pull} can be used
to show that $\kappa(h_x,h_y,\omega)$ is convex and continuous, and 
differentiable almost everywhere.  At fixed $\omega,h_x$ and $h_y$ 
$G$ converges if $z < z_c(h_x,h_y,\omega) = \exp[-\kappa(h_x,h_y,\omega)]$,
and the phase boundary $z=z_c$ is continuous.

If we turn off the interactions by setting $\omega=1$ and write
$G_0\equiv G(z;h_x,h_y,1)$ then $G_0$ satisfies the equation
\begin{equation}
G_0=(h_x+h_y)(G_0+1)z+\frac{2z^2(h_x+h_y)}{1-z}(G_0+1)
\end{equation}
so that 
\begin{equation}
1+ G_0 = \frac{1-z}{1-z-z(z+1)(h_x+h_y)}.
\end{equation}
The force-extension curve can be calculated as in section~\ref{x-pull}  and this 
has the same general form as for the two-dimensional model, and as 
found experimentally for good solvent conditions (Gunari \emph{et al.\ }
2007).

For attractive interactions ($\omega > 1$) the Temperley approach 
described in section~\ref{x-pull} for the two-dimensional case
immediately generalises to this model.  Defining partial 
generating functions as in section~\ref{model}, the partial 
generating functions obey the relations
\begin{equation}
g_0=(h_x+h_y)z(1+G)
\end{equation}
and 
\begin{equation}
g_r = (h_x+h_y)z^{r+1}\left(2+\sum_{k=0}^r (1+\omega^k)g_k +
(1+\omega^r)\sum_{k>r}g_k\right)\;,
\end{equation}
so that $g_r$ satisfies the relation
\begin{equation}
g_{r+1}-(z+q)g_r+q^r(h_x+h_y)z(q-z)g_r + qzg_{r-1}=0
\end{equation}
with $q=\omega z$.

The analysis goes through as in section~\ref{x-pull} and one can find the 
temperature dependence of the critical force.  By choosing the relative
magnitudes of $h_x$ and $h_y$ one can change the direction in the
$(x,y)$-plane in which the force is applied.  In all cases the 
force-temperature curve is monotonic and does not show 
reentrance.  Presumably this is because the model does not have 
extensive ground state entropy.  The phase diagram can also be 
calculated, using the same methods as in section~\ref{x-pull}, with qualitatively
similar results. 

\begin{figure}[ht!] 
  \psfrag{fxt}{\begin{large}$f_x^t$\end{large}}
  \psfrag{bbb}{\begin{large}$\beta^{-1}$\end{large}}
  \centering 
  \includegraphics[width=8cm]{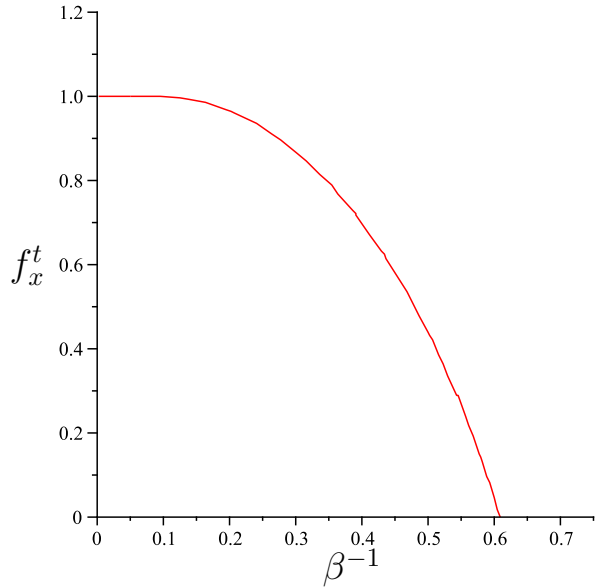}
  \caption{ 
 The temperature dependence of the critical force for the three-dimensional
model in the $n\to \infty$ limit when the force is applied in the $x$-direction.
} 
\label{3d-y1-ft} 
\end{figure} 

\begin{figure}[ht!]
  \psfrag{z}{\begin{Large}$z$\end{Large}}
  \psfrag{w}{\begin{Large}$\omega$\end{Large}}
  \centering
  \includegraphics[width=8cm]{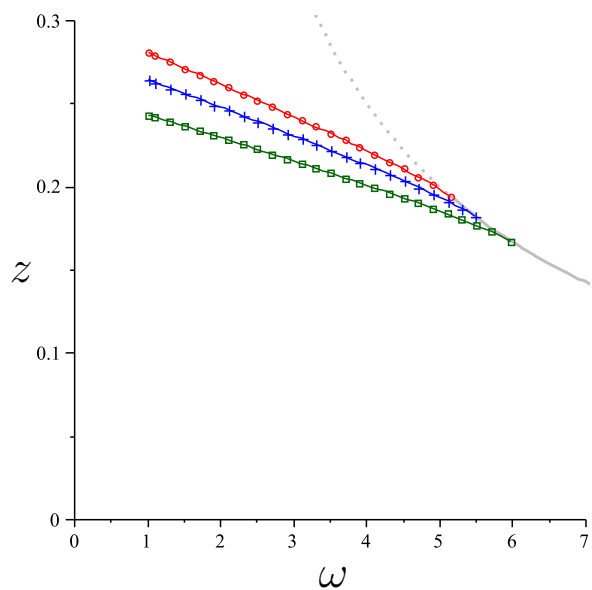}
  \caption{
The boundary of convergence of the generating function as
a function of $\omega$ for the three-dimensional model when the force
is applied in the $x$-direction.  The rectangular hyperbola is independent of
the value of $h_x$.  The three curves with points marked correspond to
$h_x=1$ (top curve), $h_x=1.2$ and $h_x=1.5$.
}
\label{3d-y1-zc}
\end{figure}

\begin{figure}[ht!]  
  \psfrag{ft}{\begin{large}$f^t$\end{large}}
  \psfrag{bbb}{\begin{large}$\beta^{-1}$\end{large}}
  \centering  
  \includegraphics[width=8cm]{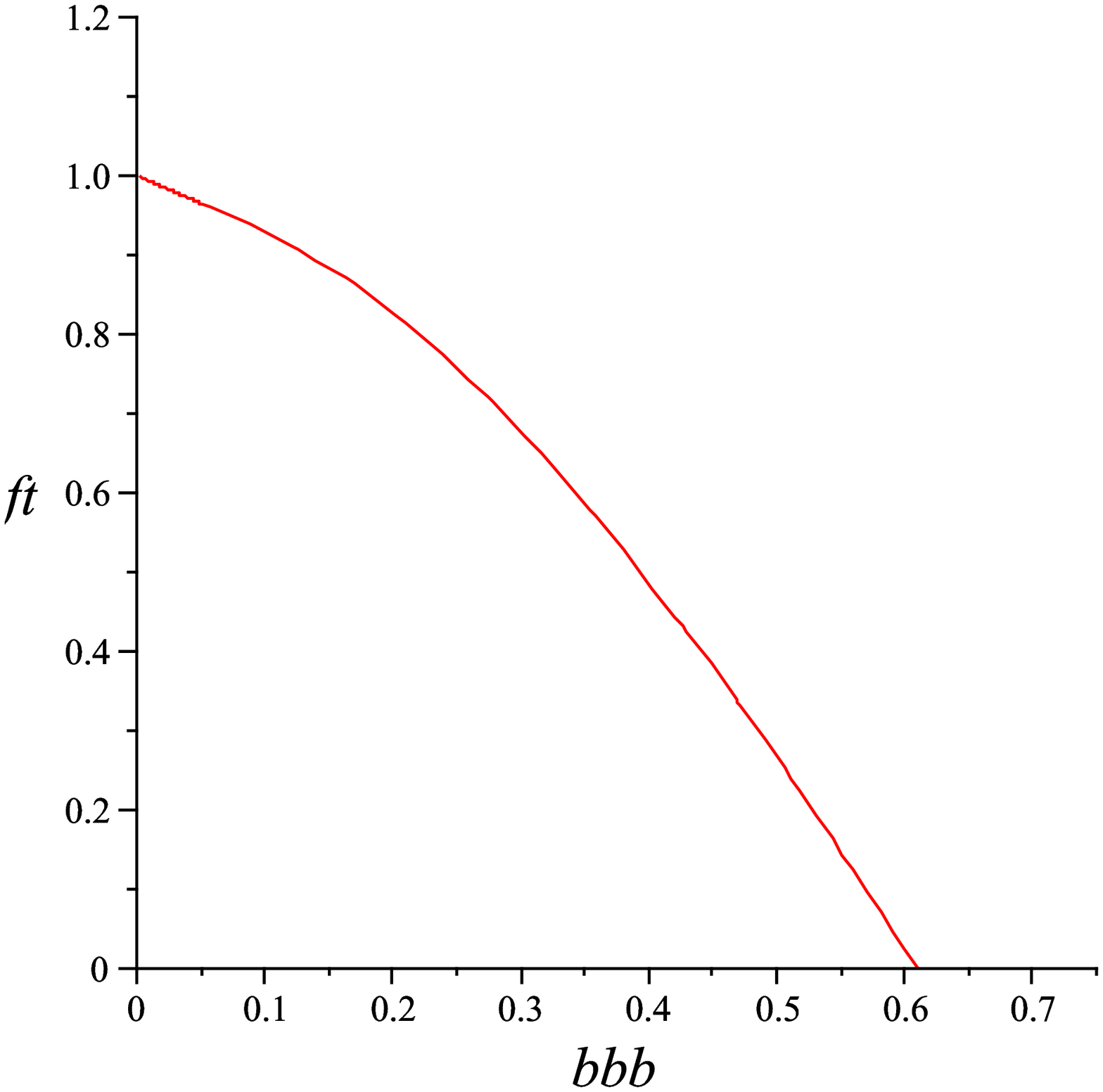}
  \caption{  
 The temperature dependence of the critical force
for the three-dimensional model in
the $n\to \infty$ limit when the force is applied in the $xy$-plane
at $45^o$ to the $x$-axis.
}  
\label{3d-yy-ft}  
\end{figure}  
 
\begin{figure}[ht!] 
  \psfrag{z}{\begin{Large}$z$\end{Large}}
  \psfrag{w}{\begin{Large}$\omega$\end{Large}}
  \centering 
  \includegraphics[width=8cm]{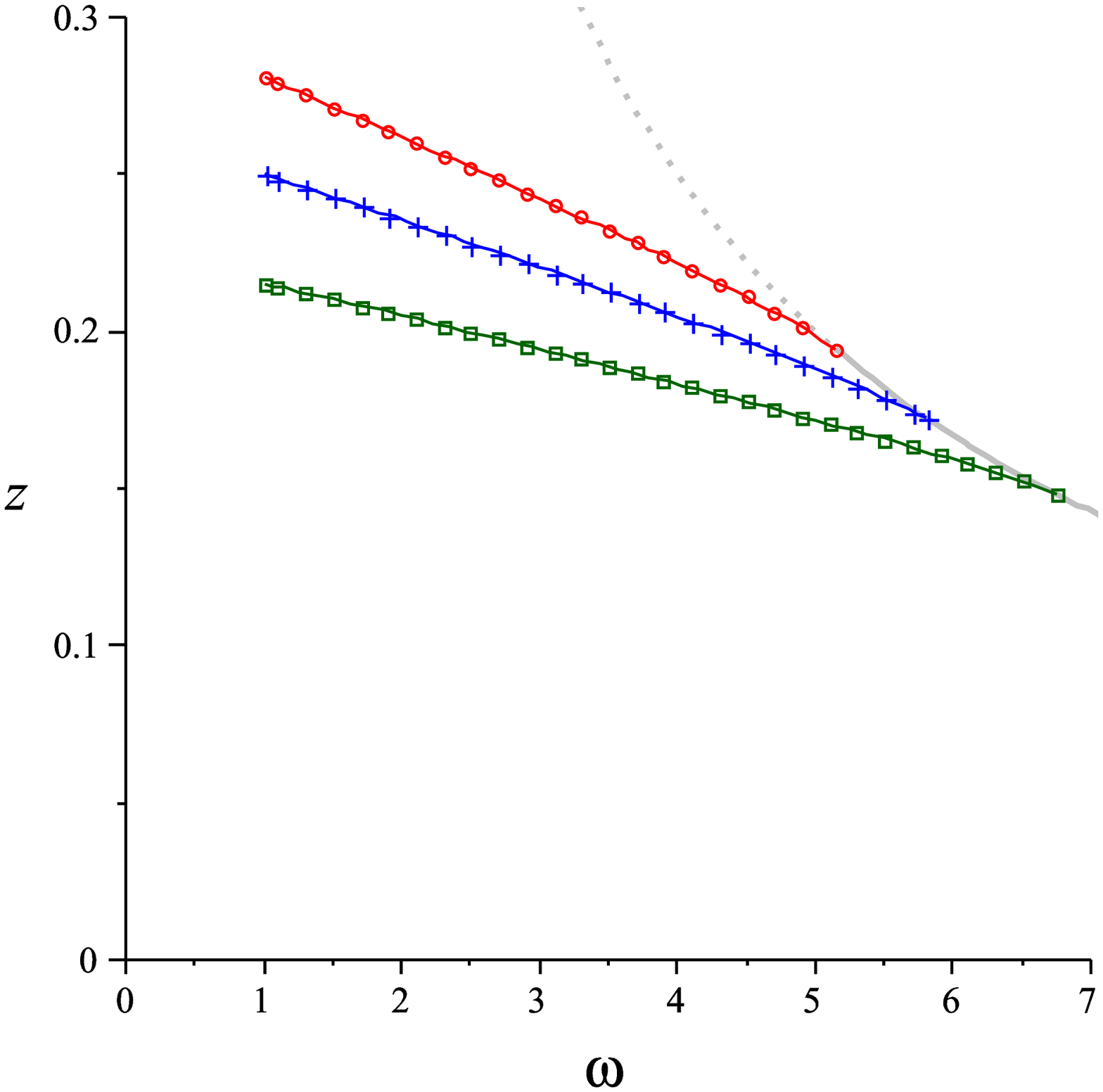}
  \caption{ 
The boundary of convergence of the generating function as
a function of $\omega$ for the three-dimensional model when the force
is applied in the $xy$-plane at $45^o$ to the $x$-axis.
The rectangular hyperbola is independent of
the value of $h_x=h_y$.  The three curves with points marked correspond to
$h_x=h_y=1$ (top curve), $h_x=h_y=1.2$ and $h_x=h_y=1.5$.
} 
\label{3d-yy-zc} 
\end{figure} 
 
If we set $h_y=0$ we reproduce the results of section~\ref{x-pull}.  We consider
two other cases.  First we set $h_x=h$ and $h_y=1$.  This corresponds 
to applying a force in the $x$-direction.  
We show the critical force-temperature curve in figure~\ref{3d-y1-ft} and the
boundary of convergence in figure~\ref{3d-y1-zc}.

These are qualitatively similar to the corresponding figures
for horizontal pulling in two dimensions, though with quantitative
differences, of course.  In particular the critical force-temperature
curve has zero slope in the $T\to 0$ limit and is monotone decreasing.
We also set $h_x=h_y=h$ which corresponds to pulling in the $xy$-plane 
but at $45^o$ to the $x$-axis.  
We show the critical
force-temperature curve in figure~\ref{3d-yy-ft} and the
boundary of convergence in figure~\ref{3d-yy-zc}.

The critical
force now has negative limiting slope in the $T\to 0$ limit but
remains monotone decreasing.   This is because of the curious feature
of this model that the ground state has no (extensive) entropy 
in the compact state but acquires entropy under the influence of a 
force.  The low temperature behaviour can be understood by the 
following crude low temperature argument.  Think of an $n$-edge walk
at low temperature $T$ under a tensile force $f$ at $45^o$ to the 
$x$-axis.  If $n-m$ edges of the walk are in a compact state 
and the remaining $m$ edges are extended the (extensive) free
energy can be written as 
\begin{equation}
F=(n-m)\epsilon -fm -Tm\log 2,
\end{equation}
where $\epsilon < 0$ is the vertex-vertex attractive energy.  
Differentiating with respect to $m$ and setting the derivative 
equal to zero gives
\begin{equation}
f=-\epsilon -T\log 2.
\end{equation}
Setting $\epsilon = -1$ gives a critical force of 1 at $T=0$ and 
the limiting slope $\mathrm{d}f/\mathrm{d}T = -\log 2 <0$.


\section{Discussion}
\setcounter{equation}{0}
\label{discussion}

We have analysed the polymer model of partially directed walks with
self-interaction, so as to induce a collapse transition, under the
influence of tensile forces on the ends of the polymer. The problem of
forces only in the preferred direction of the walk can be elucidated
completely. The phase transition which is a second-order transition without any
force is unchanged by the presence of such a force. The force
extension curves at high temperatures look qualitatively similar to
those of AFM experiments (Gunari \emph{et al.\ }(2007)).

The solution of the full model is more problematic. While the exact
solution of the generating function of partition functions can be
written down in terms of $q$-series, these functions are even more
complicated than those encountered in the standard model that has only
forces in the preferred direction (those are $q$-Bessel functions). We
have been able to solve the model on the important special curve in
parameter space which should contain the transition point. This seems
to indicate that, again, the transition is unaffected by the force.
This is more difficult to understand physically as the force must
change the high temperature state of the polymer, unlike a force
applied only in the preferred direction. It would therefore be
interesting to analyse this full model further.

\bigskip
\noindent{\bf Acknowledgements:}
Financial support from the Australian Research Council via its support
for the Centre of Excellence for Mathematics and Statistics of Complex
Systems is gratefully acknowledged by the authors. The authors would
also like to thank NSERC of Canada for financial support.

\begin{sloppypar}
\section*{References}

\vspace{0.1in} \noindent
Bemis J E, Akhremitchev B B and Walker G C 1999 {\it Langmuir}
{\bf 15} 2799--2805

\vspace{0.1in} \noindent
Binder P-M, Owczarek A L, Veal A R and Yeomans J M 1990
\emph{J. Phys. A: Math. Gen.} \textbf{23} L975--L979

\vspace{0.1in} \noindent
Bousquet-M\'elou M 1996 {\it Disc. Math.} {\bf 154} 1--25

\vspace{0.1in} \noindent
Brak R, Guttmann A J and Whittington S G 1992 {\it J. Phys. A:
Math. Gen.} {\bf 25} 2437--2446

\vspace{0.1in} \noindent
Cooke I R and Williams D R M 2003 \emph{Europhys. Lett.} \textbf{64}
267--273

\vspace{0.1in} \noindent
Grassberger P and Hsu H-P 2002 \emph{Phys. Rev. E}
\textbf{65} 031807

\vspace{0.1in} \noindent
Gunari N, Balazs A C and Walker G C 2007 \emph{J. Am. Chem. Soc.}
\textbf{129} 10046--10047

\vspace{0.1in} \noindent
Halperin A and Zhulina E B 1991 \emph{Europhys. Lett.}
\textbf{15} 417--421

\vspace{0.1in} \noindent
Haupt B J, Senden T J and Sevick E M 2002 \emph{Langmuir}
\textbf{18} 2174--2182

\vspace{0.1in} \noindent
Janse van Rensburg E J 2000 {\it Statistical Mechanics of Interacting
Walks, Polygons, Animals and Vesicles} (Oxford: Oxford University Press)

\vspace{0.1in} \noindent
Janse van Rensburg E J 2003 {\it J. Phys. A: Math.
Gen.} {\bf 36} R11--R61

\vspace{0.1in} \noindent
Krantz S G and Parks H R 2002 {\it The Implicit Function Theorem: History, Theory, and Applications}
(Birkh\"{a}user)

\vspace{0.1in} \noindent
Kumar S and Giri D 2007 \emph{Phys. Rev. Lett.}
\textbf{98} 048101

\vspace{0.1in} \noindent
{Lauritzen~Jr} J I and Zwanzig R 1970 \emph{J. Chem. Phys.}
\textbf{52} 3740

\vspace{0.1in} \noindent
Owczarek A L 1993 {\it J. Phys. A.: Math. Gen. }  {\bf 26} L647--L653

\vspace{0.1in} \noindent
Owczarek A L and Prellberg T 2007 \emph{J. Stat. Mech.} P11010

\vspace{0.1in} \noindent
Owczarek A L, Prellberg T and Brak R 1993 \emph{J. Stat. Phys.}
\textbf{72} 737--772

\vspace{0.1in} \noindent
Prellberg T 1995 \emph{J. Phys. A: Math. Gen.} {\bf 28} 1289--1304

\vspace{0.1in} \noindent
Prellberg T and Brak R 1995 \emph{J. Stat. Phys.}
\textbf{75} 701--730

\vspace{0.1in} \noindent
Prellberg T, Owczarek A L, Brak R and Guttmann A J 1992
\emph{Phys. Rev. E.}, {\bf 48}, 2386--2396

\vspace{0.1in} \noindent
Rosa A, Marenduzzo D, Maritan A and Seno F 2003 \emph{Phys. Rev. E}
\textbf{67} 141802

\vspace{0.1in} \noindent
Temperley H N V 1956 \emph{Phys. Rev.} \textbf{103} 1--16

\vspace{0.1in} \noindent
Zwanzig R and {Lauritzen~Jr} J I 1968 \emph{J. Chem. Phys.}
\textbf{48} 3351

\end{sloppypar}

\end{document}